\documentclass[letterpaper,twocolumn,pre,aps,showpacs,floatfix, 
superscriptaddress]{revtex4-2}
\usepackage[latin1]{inputenc}
\usepackage{bbm}
\usepackage{soul}
\usepackage{bm}
\usepackage[usenames]{color}
\usepackage{multirow}
\usepackage{amssymb}
\usepackage{amsbsy}
\usepackage{mathtools}
\usepackage{amsmath}
\usepackage{stmaryrd}
\usepackage{graphicx}
\usepackage{epsfig}
\usepackage{placeins}
\usepackage{bbold}
\usepackage{mathrsfs}
\usepackage{braket}
\usepackage{blindtext}
\usepackage[colorlinks,linkcolor=blue,citecolor=blue,urlcolor=blue]{hyperref}
\usepackage{physics}

\usepackage{scalerel}
\usepackage{tikz}
\usetikzlibrary{calc}
\usetikzlibrary{patterns}
\usetikzlibrary{svg.path}
\definecolor{orcidlogocol}{HTML}{A6CE39}
\tikzset{
  orcidlogo/.pic={
    \fill[orcidlogocol] 
svg{M256,128c0,70.7-57.3,128-128,128C57.3,256,0,198.7,0,128C0,57.3,57.3,0,128,
0C198.7,0,256,57.3,256,128z};
    \fill[white] svg{M86.3,186.2H70.9V79.1h15.4v48.4V186.2z}
                 
svg{M108.9,79.1h41.6c39.6,0,57,28.3,57,53.6c0,27.5-21.5,53.6-56.8,
53.6h-41.8V79.1z 
M124.3,172.4h24.5c34.9,0,42.9-26.5,
42.9-39.7c0-21.5-13.7-39.7-43.7-39.7h-23.7V172.4z}
                 
svg{M88.7,56.8c0,5.5-4.5,10.1-10.1,10.1c-5.6,0-10.1-4.6-10.1-10.1c0-5.6,4.5-10.1
,10.1-10.1C84.2,46.7,88.7,51.3,88.7,56.8z};
  }
}

\newcommand{\be}{\begin{equation}}
\newcommand{\ee}{\end{equation}}

\newcommand{\cs}{\mathcal{S}}
\newcommand{\E}{\mathcal{E}}
\newcommand{\ov}{\overline}

\newcommand{\ra}{\rangle}
\newcommand{\la}{\langle}

\newcommand{\T}{{\cal T}}

\newcommand{\HH}{{\mathcal{H}}}

\newcommand{\D}{{\mathcal{D}}}

\newcommand{\Z}{{\mathcal{Z}}}

\makeatletter

\begin{document}

\title{
Statistical structural properties of many-body chaotic eigenfunctions and applications
}

\author{Wen-ge Wang}\email{wgwang@ustc.edu.cn}
\affiliation{ Department of Modern Physics, University of Science and Technology of China,
	Hefei 230026, China}
\affiliation{CAS Key Laboratory of Microscale Magnetic Resonance,
University of Science and Technology of China, Hefei 230026, China}
\affiliation{Anhui Center for fundamental sciences in theoretical physics, Hefei 230026, China}

\author{Qingchen Li}
\affiliation{ Department of Modern Physics, University of Science and Technology of China,
	Hefei 230026, China}

\author{Jiaozi Wang}
\affiliation{Department of Mathematics/Computer Science/Physics, University of Osnabr\"uck, D-49076
Osnabr\"uck, Germany}

\author{Xiao Wang}
\affiliation{ Department of Modern Physics, University of Science and Technology of China,
	Hefei 230026, China}


\date{\today}

\begin{abstract}

In this paper, we employ a semiperturbative theory to study the statistical structural properties of energy eigenfunctions (EFs) in many-body quantum chaotic systems consisting of a central system coupled to an environment. Under certain assumptions, we derive both the average shape and the statistical fluctuations of EFs on the basis formed by the direct product of the energy eigenbases of the system and the environment. Furthermore, we apply our results to two fundamental questions: (i) the properties of the reduced density matrix of the central system in an eigenstate, and (ii) the structure of the off-diagonal smooth function within the framework of the eigenstate thermalization hypothesis. Numerical results are also presented in support of our main findings.

\end{abstract}

\maketitle

\section{Introduction}

 Among important topics that have attracted wide attentions in modern physics is the study of
intriguing behaviors of central (small) quantum systems,
 which are locally coupled to complex (huge) quantum environments
\cite{GORM,Genway13,Berman07,Berman08,Gorin14,Jin13,Olshanii14,
Jin17,Yuan11,Fialko15,Bulaev05,Wang08}.
In particular, decoherence and thermalization are two fundamental phenomena that typically appear in such generic situations.

 Due to entanglement between the central system and its environment,
 basically, properties of the total system, particularly its eigenspectrum and eigenstates,
 play a crucial role in fully understanding the behavior of the central system.
 For many-body quantum chaotic systems, regarding the eigenspectrum,
 it is now widely accepted that, its fluctuation properties can be described by the random matrix theory (RMT) \cite{Prosen181, Prosen182}.
 Meanwhile, for their eigenstates, Berry's conjecture \cite{Berry77} has been found useful for
 the description of eigenfunctions (EFs) in Billiard-type systems,
 with possible modifications due to specific dynamics \cite{KpHl,Heller87,Sr96,Sr98,Bies01,Backer02,Urb03,Kp05,Urb04,Urb06},
 and also some other type of systems \cite{Meredith88,EFchaos-Borgonovi98,EFchaos-Benet00,EFchaos-Benet03,pre18-EF-BC,PhysRevLett.134.010404-RMT-EF}.
 More importantly, inspired by the Berry's conjecture,
 the so-called eigenstate thermalization hypothesis (ETH) \cite{Deutch91,srednicki1994chaos,srednicki-JPA96,srednicki1999approach} was introduced and
  has been attracting extensive attention during the last decades.
Evidences supporting validity of ETH have been found in numerous numerical experiments \cite{Deutch-RPP18,RS-PRL12,Rigol-Aip2016}, regardless of whether the systems have classical counterparts or not,
although its theoretical foundation is still not fully understood.

However, for many-body systems, even with Berry's conjecture, analytical studies remain challenging.
 In particular, application of the correlation function given in Berry's conjecture
 involves integration over a large number of degrees of freedom,
 which is directly computable only in few special cases.
Using semiclassical theory, certain expressions for the off-diagonal function in the ETH ansatz have been derived \cite{offeth_PhysRevE.61.R2180,offeth_Wilkinson_1987,pre25-ETHsc,WW-ETH-conf}, but these expressions are generally highly involved, which restricts their practical applicability.
More recently, studies have been extended to random matrix models \cite{Nation_2018} and systems following hydrodynamic predictions \cite{PhysRevX.15.011059-Luca}. However, an explicit expression for the ETH off-diagonal function in generic systems, which is crucial for many applications of ETH, remains an open question.


 For the reasons discussed above, it is important to develop new methods for many-body chaotic systems that
 facilitate the computation of statistical structural properties of eigenfunctions,
 namely, their average shape and statistics of fluctuations, and thereby be useful
 in the study of thermalization and decoherence properties of the central system.

The main purpose of this paper is to derive the statistical structural properties of eigenfunctions (EFs)
of the total system in a basis formed by the direct product of the energy bases of the central system and its environment.
We refer to this as the \textit{unperturbed energy basis}, and treat the system-environment interaction as a perturbation.
The main ingredient used here is a semi-perturbation theory introduced in
Ref.~\cite{WIC98,pre00-GBW, pre02-GBW, CPL04,CPL05,JPA19-NPT-C,CTP19-renorm}.
Once these properties are known, tracing out the environmental degrees of freedom would correspond to
integration or summation over only one variable --- the environmental energy.

 The paper mainly consists of three parts.
 In the first part, a framework is established for the purpose mentioned above.
 The strategy is to start from a generic many-body model of physical relevance
 and, then, introduce a series of assumptions for restriction step by step,
 until a specific model is finally reached.
 In the second part, we are to derive explicit expressions for statistical structural properties of EFs
 in the this model.

 In the third part, we discuss some applications of results of the second part.
The first application concerns the foundation of the \textit{eigenstate thermalization hypothesis} (ETH),
which remains not fully understood to date.
 The second application concerns the possibility of deriving a Gibbs form for the reduced density matrix (RDM) of the central system, obtained by tracing out the environmental degrees of freedom. Interestingly, this study also suggests the occurrence of \textit{eigenstate decoherence} in the central system, with its energy basis serving as a preferred (pointer) basis.
\footnote{Note that the ETH ansatz does not guarantee eigenstate decoherence, nor does it guarantee a Gibbs form of the RDM of the central system
  \cite{pre20-mc-can}}



For a more detailed guide, the rest of the paper is organized as follows.
Preliminary discussions are presented in Sec.~\ref{framework}.
In Sec.~\ref{sect-model-m-results}, we discuss the problems to be addressed and the systems under study.
The basic framework of the main analytical approach, the semiperturbative theory, is discussed in Sec.~\ref{sect-semipertabative}.
The central part of the paper is contained in Secs.~\ref{sect-Ak} and \ref{sect-structural-properties-EFs}, where we derive expressions for high-order perturbation contributions to EFs (Sec.~\ref{sect-Ak}) and analyze the statistical structural properties of EFs on the unperturbed energy basis (Sec.~\ref{sect-structural-properties-EFs}).
Applications of these results, including those to the two questions discussed above, are presented in Sec.~\ref{sect-app}.
Numerical simulations for illustrating the analytical predictions are given in Sec.\ref{sect-numerics}. 
Finally, conclusions and discussions are provided in Sec.~\ref{sect-conclusion}.

\section{Preliminary discussions}\label{framework}

 In this section, we discuss basic notations for systems under study (Sec.\ref{sect-notation})
 and introduce basic assumptions of restriction (AR),
 which are useful for setting up a framework
to study the thermalization of a central system.
 There are totally six ARs, which will be referred to as AR1, AR2, ...., AR6.
 The first one is introduced in Sec.\ref{sect-notation}
 and the others in Secs.\ref{AR2-6}.

\subsection{Basic notations}\label{sect-notation}

Let us consider a quantum many-body system, denoted by $\T$.
 It is divided into a central small system, denoted by $S$, and a large environment,
 denoted by $\E$.
 Numbers of particles (or qubits, etc.) in $\T, S$, and $\E$ are denoted by $N_\T$,
 $N_S$, and $N_{\E}$, respectively, with $N_S \ll N_{\E}$.
 We use $\HH_S$ to indicate the Hilbert space of the central system $S$, with a dimension $d_S$;
 and $\HH_{\E}$ for the Hilbert space of the environment,
 with a dimension $d_{\E}$.
 For the sake of simplicity in discussion, we assume that the central system $S$ has a nondegenerate spectrum.

 The Hamiltonians of $S$ and ${\E}$, which are determined in the weak-coupling limit,
 are indicated as $H^{S}_0$ and $H^{{\E} }_0$, respectively,
 and the $S$-${\E}$ interaction Hamiltonian is indicated as $ H^{I}_0$.
 The total Hamiltonian $H$ is then written as
\begin{equation}\label{H-total}
 H = H^{S}_0 + H^{{\E} }_0 +  H^{I}_0.
\end{equation}
 Eigenstates of the total system are denoted by $|n\ra$ with energies $E_n$
 in the increasing-energy order,
\begin{gather}\label{Seq-H}
 H|n\rangle = E_n|n\rangle.
\end{gather}
 One remark: The operator $H^{S}_0$ on the right-hand side (rhs) of Eq.(\ref{H-total}) should,
 more exactly, be written as
 $H^{S}_0 \otimes I^{\E}$, where $I^{\E}$ represents the identity operator on the Hilbert space of ${\E}$.
 For brevity, we usually omit such an identity operator, unless we want to stress it;
 and similar for $H^{{\E} }_0$.

 The first assumption, AR1, is about locality of the interaction.
\begin{itemize}
  \item AR1. The interaction between the central system $S$ and the environment $\E$ is local.
\end{itemize}
 This assumption implies that the environment ${\E}$ may be further divided into two parts:
 A small part to be indicated as $A$ with $N_A$ particles
 and a large part indicated as $B$ with $N_{B}$ particles ($N_{B} \gg N_A$),
 such that only the $A$ part is coupled to the central system $S$.
 Hilbert spaces of the part $A$ and the part $B$
 are to be indicated as $\HH_A$ and $\HH_{B}$,
 respectively, with dimensions $d_A$ and $d_{B}$.
 Clearly, $d_{\E} = d_A d_{B}$.
 The operator $H^{I}_0$ is assumed to have a product form,
\begin{equation}\label{HI}
H^{I}_0=  H^{IS}\otimes H_0^{IA},
\end{equation}
 where  $H^{IS}$ and $H_0^{IA}$ are Hermitian operators
 that act on the Hilbert spaces of $S$ and $A$, respectively.

 The interaction may have some average effect on the central system.
 In this case, it is usually convenient to move the average effect
 to the self-Hamiltonian part of the central system.
 For this purpose, one may write the total Hamiltonian in an equivalent
 form of $H= H^S + H^I + H^{\E}$, where
\begin{subequations}\label{}
\begin{align}\label{}
 & H^S = H^{S}_0 +  H^{IS} h_0,
 \\ &  H^I = H^{IS}\otimes H^{IA}, \label{HI-OS-h0}
 \\ &  H^{\E} = H^{{\E} }_0,
\end{align}
\end{subequations}
 Here, $h_0$ is a c-number, representing certain average effect of $H_0^{IA}$,
 \footnote{The exact averaging method for the computation of $h_0$
 may depend on the problem at hand.
 For example, when only a certain energy region is of relevance to the problem,
 the average may be taken within the energy region, not over the whole energy spectrum.}
 and
\begin{align}\label{}
 &  H^{IA} = H_0^{IA} -  h_0. \label{HIA}
\end{align}

 Normalized eigenstates of $H^S$ are denoted by $|\alpha\rangle$ with energies $e^S_\alpha$
 and those of $H^{\E}$ by $|i\rangle$ with energies $e_i^{\E}$,
 both in the increasing-energy order;
\begin{subequations}\label{Sa-Ei}
\begin{align}\label{}
\label{Sa}
 & H^S|\alpha\rangle = e^S_\alpha|\alpha\rangle, &   \alpha =1,\ \ldots, d_S;
\\ & H^{\E}|i\rangle = e^{\E}_i|i\rangle, &   i =1,\ \ldots, d_{\E_0}. \label{Ei}
\end{align}
\end{subequations}
 It is sometimes convenient to indicate the interaction strength in an explicit way
 and, in this case,
 we write
\begin{align}\label{}
 H^I \equiv \lambda H^I_{\rm norm},
\end{align}
 where $\lambda$ is a parameter
 for characterizing the interaction strength and $H^I_{\rm norm}$
 indicates  some normalized version of the operator $H^I$.

In the discussions in later sections, the interaction Hamiltonian $H^I$ is regarded as a perturbation, and the total Hamiltonian reads
\begin{equation}\label{H=H0+HI}
 H =  H^0 + H^{I},
\end{equation}
 where $H^0$ is the unperturbed Hamiltonian,
\begin{equation}\label{H0}
 H^0 := H^{S} + H^{{\E}}.
\end{equation}
 The basis given by the uncoupled states of $\{|\alpha\rangle |i\rangle \}$, in short $\{|\alpha i \rangle \}$,
 is the \textit{unperturbed energy basis} mentioned previously.
 Ordered in energy, the states $|\alpha i\ra$ are to be indicated as $|E_r^0\ra$,
 with a label $r$ in a one-to-one correspondence to the pair $(\alpha, i)$, i.e.,
 \begin{gather}\label{r-alphai}
 r\leftrightarrow (\alpha, i), \quad r=1,2,3, \ldots.
\end{gather}
 That is, $|E_r^0\ra$ are the unperturbed basis states with unperturbed energies $E_r^0$,
\begin{align}\label{Seq-H0}
 & H^0 |E_r^0\ra = E_r^0|E_r^0\ra,
 \\ &  E_r^0 = e^S_\alpha + e^{\E}_i \quad (E_r^0 \le E_{r+1}^0).
\end{align}

 For the simplicity in discussion, we assume time reversal symmetry of the total Hamiltonian
 such that its elements are real.
 Expansions of the total eigenstates $|n\ra$ on the bases of $|\alpha i\ra$ and of
 $|E_r^0\ra$ are written as
 \begin{equation}\label{|n>}
 |n \rangle = \sum_{\alpha i}C_{\alpha i}^n |\alpha i\rangle = \sum_{r}C_{r}^n |E_r^0\rangle;
\end{equation}
 and, they give EFs of $C_{\alpha i}^n$ and $C_r^n$, respectively.

 Elements of $H^I$ on the above basis are indicated as $H^{I}_{rs} $,
\begin{align}\label{}
 H^{I}_{rs} \equiv H^I_{\alpha i,\beta j}  = \la \alpha i|H^I|\beta j\ra ,
\end{align}
 with $r \leftrightarrow (\alpha, i)$ and $ s \leftrightarrow (\beta, j)$.
 From Eq.(\ref{HI-OS-h0}), one gets that
\begin{align}\label{HIrs=HISab-HIAij}
 H^{I}_{rs} = H^{IS}_{\alpha\beta} H^{IA}_{ij},
\end{align}
 where
\begin{subequations}
\begin{gather}
 H^{IS}_{\alpha\beta} = \la \alpha |H_0^{IS}|\beta\ra,
\\  H^{IA}_{ij} = \la i |H_0^{IA}|j\ra - h_0 \delta_{ij} . \label{HIA-ij}
\end{gather}
\end{subequations}

\subsection{Assumptions of AR2-AR6}\label{AR2-6}

 The second assumption of restriction is about a banded structure of the interaction Hamiltonian.
\begin{itemize}
  \item AR2.  The matrix of $H^{IA}_{ij}$ possesses a banded structure,
 with a band width indicated by $b^{H}_{IA}$, that is,
\begin{align}\label{HIAij-band}
 H^{IA}_{ij} =0 \quad \text{if $|e^{\E}_i - e^{\E}_j| > b^{H}_{IA}$}.
\end{align}
\end{itemize}
 One may note that AR2 is (approximately) satisfied in most physical many-body models with local interaction,
 particularly when the energy scale of the $A$ part of the environment is much smaller
 than the $B$ part.

 A further restriction is that the system-environment interaction is not very strong,
 such that main bodies of the EFs $C^n_r$ are narrow
 compared with the band width $b^{H}_{IA}$.
 (See Appendix \ref{sect-main-body-region} for a mathematical definition of the main body of an EF.)
 We use $w_{\rm mb}^{\rm max}$ to indicate the maximum width of the main-body regions of all
 the EFs of interest.
 The third assumption of restriction is written as follows.
\begin{itemize}
  \item AR3. $\lambda < \lambda_{\rm UB}$,
 where $\lambda_{\rm UB}$ is an upper bound of the interaction strength for which
 $w_{\rm mb}^{\rm max} \ll b^{H}_{IA}$.
\end{itemize}

 The fourth assumption of restriction
 is about a lower bound of the system-environment interaction strength,  denoted by $\lambda_{\rm LB}$.
\begin{itemize}
  \item AR4. $\lambda > \lambda_{\rm LB}$.
\end{itemize}
 Such a lower bound is needed in the study of thermalization.
 \footnote{Explicit estimates to $\lambda_{\rm UB}$ and $\lambda_{\rm LB}$
 for the systems to be studied will be given in Sec.\ref{sect-main-results}.
 }
 In fact, as a common sense
 which is based on both intuitive physical pictures and simple perturbation-theory arguments,
 thermalization is not expected to happen when the interaction is sufficiently weak.
 More exactly, the interaction at $\lambda_{\rm LB}$
 should possess a strength that is much larger than the mean level spacing of the environment,
 namely, much larger than $1/\rho^{\E}_{\rm dos}$,
 where $\rho^{\E}_{\rm dos}$ indicates the density of states of the environment.

 In the study of thermalization, it is necessary to assume certain complexity,
 more exactly, certain random properties of the environment.
 These are contents of the fifth and sixth assumptions of restriction.

The fifth assumption of restriction, AR5, is about the distribution of
 the elements  $H^{IA}_{ij}$ of the environmental part of the interaction Hamiltonian.
 Due to AR2, we need to consider only those elements $H^{IA}_{ij}$ within the coupling band,
 for which $|e^{\E}_i - e^{\E}_j| \le  b^{H}_{IA}$.
\begin{itemize}
  \item AR5.
  The distribution of $H_{ij}^{IA}
  [\rho_{{\rm dos}}^{{\cal E}}(e_{i}^{{\cal E}})\rho_{{\rm dos}}^{{\cal E}}(e_{j}^{{\cal E}})]^{1/4}$
 for $|e^{\E}_i - e^{\E}_j| < b^{H}_{IA}$
 possesses a mean zero and a variance denoted by $\sigma_{IA}^2$.
\end{itemize}
 It should be noted that AR5, particularly the term of
 $ [\rho_{{\rm dos}}^{{\cal E}}(e_{i}^{{\cal E}})\rho_{{\rm dos}}^{{\cal E}}(e_{j}^{{\cal E}})]^{1/4}$,
is consistent with numerical simulations reported in various models in the literature, particularly those performed for checking validity of ETH.
 In the special case that $\rho^{\cal E}_{\text{dos}}$ may be approximated as a constant
 within $b^H_{IA}$, an estimate to $\sigma_{IA}^2$ is possible, which gives that
 (see Appendix \ref{app-prove-sigmaIA})
\begin{align}\label{sigmaIA-estim}
   \sigma_{IA}^2  \simeq  \frac{1}{ 2b^{H}_{IA} } \frac{1}{d_A }  \tr_A (H^{IA})^2.
\end{align}

 The assumption AR6 is about a scaling behavior of the sum of
 products of those elements $H^{IA}_{ij}$ that lie within the above discussed band structure;
 and, only those products are considered, the
 elements in each which are not directly correlated in the sense of
\begin{align}
 & H^{IA}_{i j} \ne H^{IA}_{i' j' } \ \ \& \ \
 H^{IA}_{ij} \ne (H^{IA}_{i' j'})^*. 
 \label{gamma-require}
\end{align}
 Concretely, we use $\gamma$ to indicate a sequence of $(L+1)$ labels $i_l$,
\begin{align}
 \label{gamma}
  \gamma & =  \{ i_l:  l=0,1, \ldots ,L \},
\end{align}
 for which (i)  $|e^{\E_0}_{i_l} - e^{\E_0}_{i_{l'}}| < b^{H}_{IA}$ for all $l$ and $l'$,
 and (ii) all the elements $H^{IA}_{i_l i_{l+1}}$ of $l<L$ satisfy Eq.(\ref{gamma-require}).
 And, we use $\gamma_{\rm sub}$ to indicate a subset of $\gamma$, which contains neither $i_0$
 nor $i_L$, i.e.,
\begin{align} \label{gamma-sub}
  \gamma_{\rm sub} & =  \{ i_l \quad \text{of a number of $l \ne 0,L$} \}.
\end{align}
 AR6 concerns the following sum indicated as $Q_{\gamma_{\rm sub}}$,
\begin{align}\label{QM}
 & Q_{\gamma_{\rm sub}}= \sum_{i_l \in \gamma_{\rm sub}}  \prod_{l=0}^{L-1} H^{IA}_{i_l i_{l+1}}.
\end{align}
\begin{itemize}
  \item AR6. The quantity $Q_{\gamma_{\rm sub}}$ scales as
\begin{align}\label{AR6}
  Q_{\gamma_{\rm sub}} \sim \sqrt{\sum_{i_l \in \gamma_{\rm sub}} 1}  \qquad \forall  \gamma_{\rm sub}.
\end{align}
\end{itemize}

It is worth noting that, AR6 does not assume vanishing correlations between the matrix elements of the interaction Hamiltonian and, hence, is considerably weaker than the ETH ansatz.


\section{The model and main results}\label{sect-model-m-results}

 In this section, based on discussions in the previous section,
 we introduce the model to be studied (Sec.\ref{sect-model}).
 We then present the main results concerning statistical structural properties of
 the EFs of the model (Sec.\ref{sect-main-results}).

\subsection{The model to be studied}\label{sect-model}

 Suppose one is interested in an arbitrarily chosen  eigenstate of the total system indicated as $|n_0\ra$.
To study main properties of this state, let us recall
 AR3, i.e., $w_{\rm mb}^{\rm max} \ll b^{H}_{IA}$.
 It is then sufficient to consider the subspace spanned by the basis states $|E^0_r\ra$, whose energy lies within
 the bandwidth, i.e.,
\begin{align}\label{E0r-band}
 E^0_r \in (E_{n_0} - \frac 12 b^{H}_{IA}, E_{n_0} + \frac 12 b^{H}_{IA} ).
\end{align}
This is the underlying idea behind the model we propose and study below.
 For the simplicity in discussion, we assume that the central system's energy scale
 is much smaller than the band width $b^{H}_{IA}$, i.e.,
\begin{align}\label{}
 e^S_{d_S} - e^S_1 \ll b^{H}_{IA}.
\end{align}

 Properties of the central small system are to be retained,
 while, the environmental Hilbert space is to be truncated.
 {The ``truncated'' environment may be denoted by $\E_{\text{tr}}$.
 However, for brevity, we are to omit the subscript ``$\text{tr}$''
 in the remainder of the paper and refer to $\E_{\text{tr}}$ simply as $\E $.}

 More exactly, for the purpose of studying $|{n_0}\ra$,
 the environmental spectrum is truncated to the energy shell of
 $ (e^\E_{\rm LB}, e^\E_{\rm UB})$, where
 \begin{subequations}\label{e-E-UB-LB}
  \begin{align}
 & e^\E_{\rm LB} = E_{n_0}  - \frac{e^S_{d_S} + e^S_1}{2} - \frac 12 b^{H}_{IA},
 \\ & e^\E_{\rm UB} = E_{n_0}  - \frac{e^S_{d_S} + e^S_1}{2} + \frac 12 b^{H}_{IA}.
  \end{align}
 \end{subequations}
 The truncated subspace is to be indicated as $\mathcal{H}_\E$,
 with a superscript for ``truncated'' omitted for brevity,
\begin{align}\label{}
 \HH_\E = \bigoplus_{i} |i\ra \quad \text{for $e^\E_i \in (e^\E_{\rm LB}, e^\E_{\rm UB})$}.
\end{align}
 Its dimension is indicated as $d_\E$, which is still far much larger than $d_S$.
 Note that, with Eq.(\ref{e-E-UB-LB}), the assumption of AR3 which means
 that $w_{\rm mb}^{\rm max} \ll b^{H}_{IA}$ is written in the following form.
\begin{align}\label{}
 & \text{AR3: $\lambda < \lambda_{\rm UB}$}   \Longleftrightarrow
 w_{\rm mb}^{\rm max} \ll (e^\E_{\rm UB} - e^\E_{\rm LB}).
\end{align}

 Now, we are ready to introduce the model to be studied in later sections.
 The model consists of a central small system, which is just the system $S$ as discussed previously,
 and an environment, whose Hilbert space is $\HH_\E$.
 The total Hamiltonian $H$ is also written in the form of Eq.(\ref{H=H0+HI}), i.e.,
 $H =  H^0 + H^{I}$,
 and eigenstates of $H$ is also denoted by $|n\ra$.
 The interaction Hamiltonian $H^I$ is also written in the product form of $H^I = H^{IS}\otimes H^{IA}$
 [Eq.(\ref{HI-OS-h0})].
 Here, $H^{IA}$ is defined by its elements $H^{IA}_{ij}$ on environmental states in $\HH_\E$.

Regarding the assumptions of restriction, the effects of AR1 and AR2 have been clearly already incorporated into the model.
 The two assumptions of AR3 and AR4 are written as
 $\lambda_{\rm LB} < \lambda < \lambda_{\rm UB}$.
 Concerning AR5 and AR6, one may note that the matrix $[H^{IA}_{ij}]$
 is full in this model and, as a consequence, the condition of $|e^\E_i - e^\E_j| < b^{H}_{IA}$
 is now automatically satisfied, which has been used in AR5
 and in the first part of the definition of sequence $\gamma$ in Eq.(\ref{gamma}).

 The above arrangements of the model guarantees that structural properties of its EFs in the middle energy
 region should be similar to those of $|n_0\ra$ in the original total model.
 Hence, instead of studying the latter, one may study the former.
 Moreover, the model is of relevance by itself, though not as generic as the original model.

 Finally, making use of Eq.(\ref{sigmaIA-estim}), it is easy to see
 that $|H^{IA}_{ij}|^2$ should scale qualitatively in a way similar to the mean level spacing,
 which is very small for a many-body system.
 As a result,  the elements $H^{IA}_{ij}$ are typically much larger than the  mean level spacing.
 This implies that the interaction is usually strong, except for
 the case of extremely small $\lambda$.

\subsection{Main results}\label{sect-main-results}

 Main results to be derived in later sections are about statistical structural properties of EFs
 for states $|n\ra$ in the middle energy region.
 Specifically, they concerns division of each EF into two parts,
 which are to be referred to as nonperturbative (NPT) and perturbative (PT) parts.
 \footnote{Exact definition of the division is to be given in Sec.\ref{sect-SPT-NPT}.}

 The NPT part includes those components $C^n_r = \la E_r^0|n\ra$,
 for which the basis states $|E_r^0\ra$ belong to a set $\ov S_n^{\rm NPT}$ defined by
\begin{gather}\label{ovSn-r1r2-NPT}
 \ov S_n^{\rm NPT} := \left\{|E_r^0\ra :  r_1^{\rm NPT} \le r \le r_2^{\rm NPT} \right\},
\end{gather}
 with energies around the exact eigenenergy $E_n$,
\begin{align}\label{}
 & E_{r_2^{\rm NPT}}^0 - E_n \simeq E_n - E_{r_1^{\rm NPT}}^0.
\end{align}
 The energy width of the NPT set is denoted by $w_{\rm NPT}$,
\begin{align}\label{}
 & w_{\rm NPT} = E_{r_2^{\rm NPT}}^0 - E_{r_1^{\rm NPT}}^0.
\end{align}
 The PT part of the EF is on the rest basis states, $|E_r^0\ra \in S_n^{\rm PT}$,
 where $S_n^{\rm PT}$ is the complementary set of $\ov S_n^{\rm NPT}$.

 There are two main results.
 The first one involves two quantities:
 $\xi_{\rm w}$, which is the maximum eigenvalue of a $d_S \times d_S$ matrix $m_{\rm w}$
 defined by $(m_{\rm w})_{\alpha \alpha'} =|H^{IS}_{\alpha \alpha'}|^2$,
 and $\xi_{\rm s}$, which is the maximum (absolute) eigenvalue of $H^{IS}$.
 For example, in the special case
 that all the elements of $H^{IS}$ have a same value denoted $h_S (>0)$,
 one has $\xi_{\rm w} \simeq  d_S h_S^{2}$ and $\xi_{\rm s} \simeq  d_S h_S $.
 The first main result is the following estimation to the width of the NPT part,
\begin{gather}\label{w-NPT}
 w_{\rm NPT} \simeq 4  \sigma_{IA}^2 \xi_{\rm w} (\text{or $\xi_{\rm s}^2$}).
\end{gather}
 Here, $\xi_{\rm w}$ works for the case of weak interactions,
 which satisfies the requirement that
 $w_{\rm NPT} \ll |e_{\beta}^S-e_{\alpha}^S|$ for all $\alpha \ne \beta$;
 while, $\xi_{\rm s}$ is for the case of relatively strong interactions, such that
 $w_{\rm NPT} \gg |e_{\beta}^S-e_{\alpha}^S|$ for all $\alpha \ne \beta$.
 [In the generic case, $w_{\rm NPT}$ satisfies the relation to be given in Eq.(\ref{eta-xi=1}).]

 In fact, the width $w_{\rm NPT}$ provides a measure of the interaction strength on the
 unperturbed energy basis and it can be used to give a quantitative expression for AR4.
 That is,
\begin{align}
 & \text{AR4 of $\lambda > \lambda_{\rm LB}$}
 \Longleftrightarrow {w_{\rm NPT}} \gg \frac{1}{\rho^\E_{\rm dos}}.
\label{Lam-LB}
\end{align}

 The second main result is an expression for PT parts of the EFs,
 which works in the whole interaction region of $\lambda_{\rm LB} < \lambda < \lambda_{\rm UB}$.
 That is,
\begin{align}\label{nr-PT}
    & C^n_r \simeq  \frac{J^n_r}{E_n -E^0_{r}} \qquad \text{for $|E^0_r\ra \in S_n^{\rm PT}$ },
\end{align}
 where $J_r$ have random phases in the sense of AR6.
 When the environmental density of states is approximated as a constant, namely,
 its average value indicated as $\ov\rho^\E_{\rm dos}$, local average of $|J^n_r|^2$
 does not depend on the label $r$ and is approximately given by
\be\label{eq-Jr2}
\overline{|J^n_{r}|^{2}} \simeq
\frac{1}{\ov\rho^\E_{\rm dos}} \omega_{\text{NPT}}\left(1-W_{n}^{\rm NPT}\right).
\ee
 Here, $W_n^{\rm NPT}$ represents the weight of the NPT part of the EF,
\begin{align}\label{Wn-NPT-weight}
 & W_n^{\rm NPT} := \sum_{|E^0_{s} \ra \in  \ov{S}_n^{\rm NPT}} |C^n_s|^2.
\end{align}
 Note that, making use of the two main results,
 one may get an estimate to $w_{\rm mb}^{\rm max}$.

 To derive the above discussed two main results,
 we are to make use of a so-called semiperturbative theory, which will be explained in the next section.
 The derivations will then be presented in Secs.\ref{sect-Ak} and \ref{sect-structural-properties-EFs}.

\section{Semiperturbative theory}\label{sect-semipertabative}

 In this section, we recall basic contents of the semiperturbative theory
 \cite{WIC98,pre00-GBW,pre02-GBW,CTP19-renorm,JPA19-NPT-C}.
 Within the framework of this theory, each EF is divided into a NPT part and a PT part,
 such that components in the PT part can be expanded in a convergent perturbation
 expansion in terms of those in the NPT part.
 This expansion, known as a generalized Brillouin-Wigner perturbation expansion (GBWPE)
 \cite{WIC98}, is discussed in Sec.\ref{sect-SPT-GBWPE}.
 Then, the division into PT and NPT parts is discussed in Sec.\ref{sect-SPT-NPT}.
 Finally, an extension of the semiperturbative theory
 will be given in Sec.\ref{sect-ST-error}, which tolerates a finite error.

\subsection{The GBWPE}\label{sect-SPT-GBWPE}

 In a semiperturbative approach to the EF of a state $|n\ra$ on the basis of $|E_r^0\rangle $,
 the set $\{ |E_r^0\rangle \}$ is divided into two subsets, denoted by $S_n$ and $\ov S_n$.
 Correspondingly, the total state space is divided into two subspaces
 denoted by $\HH_{S_n}$ and $\HH_{\ov S_n}$,
 respectively, which are spanned by $|E_r^0\ra \in S_n$ and $|E_r^0\ra \in \ov S_n$,
 with projection operators denoted by $P_{S_n}$ and $Q_{\ov S_n}$,
\begin{subequations} \label{PQ}
\begin{gather}\label{}
P_{S_n} = \sum\limits _{|E_r^0\rangle\in S_n}|E_r^0\rangle\langle E_r^0|, \
\\ Q_{\ov S_n} = \sum\limits _{|E_r^0\rangle \in {\ov S_n}}|E_r^0\rangle\langle E_r^0|.
\end{gather}
\end{subequations}
 Clearly, $P_{S_n} + Q_{\ov S_n} =1$ and $P_{S_n} H^0 = H^0 P_{S_n}$.
 The  state $|n \ra$ is then divided into two parts,
\begin{gather}\label{eq-apq}
 |n  \ra = |n_P\ra + |n_Q\ra,
\end{gather}
 where
\begin{subequations} \label{|n>PQ}
\begin{gather}\label{}
 |n_{P}\rangle = {P_{{S_n}}|n \rangle},
 \\ |n_{Q}\rangle = Q_{{\ov S_n}}|n \rangle.
\end{gather}
\end{subequations}

 Multiplying both sides of Eq.~\eqref{Seq-H} by $P_{S_n}$, one gets that
 $(E_n  -H^0)|n_P\ra =  P_{S_n} H^I |n \ra$.
 This gives
\begin{gather}\label{alpha-P-1}
|n_{P}\rangle=T_n|n \rangle,
\end{gather}
 where $T_n$ is an operator defined by
\begin{equation}\label{T-alpha}
 T_n :=  \frac{1}{E_n -H^{0}} P_{S_n} {H^I}.
\end{equation}
 Substituting Eq.\eqref{eq-apq} into the rhs Eq.\eqref{alpha-P-1}, one gets that
\begin{gather}\label{nP-nQ}
 |n_{P}\rangle=T_n|n_{Q}\rangle +T_n|n_{P}\rangle.
\end{gather}
 By iterating Eq.~\eqref{nP-nQ}, one gets the following expansion,
\begin{gather}
|n_{P}\rangle= \sum_{l=1}^{k-1} \left( T_n \right)^{l} |n_Q  \rangle
 + \left( T_n \right)^k |n_P  \rangle.
\label{nP-expan}
\end{gather}

 Practically, the state $|n \ra$ is usually unknown and, as a result,
 it is impossible to compute the exact vector of $\left( T_n \right)^k |n_P\ra $.
 Instead, one may consider an arbitrary normalized vector $|\psi^{(0)} \ra$ in the subspace $\HH_{S_n}$
 and its $k$-th order weight, denoted by $A_k$,
\begin{align}\label{Ak}
 &  A_k := \la \psi^{(k)}|\psi^{(k)}\ra,
\end{align}
 where
\begin{align} \label{psi-(k)}
 & |\psi^{(k)}\ra : = (T_n)^k |\psi^{(0)} \ra .
\end{align}
 Clearly,  one gets the following convergent perturbation expansion of $|n_P\ra$,
\begin{align}\label{nP-expan-infty}
 |n_{P}\ra = \sum_{l=1}^{\infty} \left( T_n \right)^{l} |n_Q  \rangle,
\end{align}
 if
\begin{align}\label{conv}
 \lim_{k \to \infty} A_k =0 \quad \forall \, |\psi^{(0)}\ra \in \HH_{S_n}.
\end{align}
 The rhs of Eq.~\eqref{nP-expan-infty} is the GBWPE
 of the $|n_{P}\ra$-part of the  state $|n\ra$.

 A further simplification to Eq.~\eqref{conv} is possible.
 To this end, let us write $|\psi^{(0)}\ra $ as
\begin{align}\label{}
 |\psi^{(0)}\ra = {\sum_{s_0}}' c_{s_0} |E^0_{s_0}\ra,
\end{align}
 with ${\sum_{s_0}}' |c_{s_0}|^2 =1$.
 Here and hereafter, the prime over a summation symbol ``$\sum_r$''
 means that the summation is over $|E^0_r\ra \in S_n$, i.e.,
\begin{gather}\label{prime-sum-r}
 {\sum_r}' := \sum_{r: |E_r^0\ra \in S_n}.
\end{gather}

 The quantity $A_k$ is then written as
\begin{align}\label{Ak=sum-rs-phi}
 A_k =  {\sum_{{s_0},s_0'}}' c_{s_0'}^* c_{s_0}  \la E^0_{s_0'}|(T_n^\dag)^k(T_n)^k |E^0_{s_0}\ra .
\end{align}
Clearly, in the limit of $k\to \infty$, if $\| (T_n)^k |E^0_{s_0}\ra \|$
vanishes for all $|E^0_{{s_0}}\ra \in S_n$, then, $A_k \to 0$.
 Hence, the condition in Eq.\eqref{conv} for the GBWPE is equivalently written as the following condition,
\footnote{Formally, one may consider an operator $W_S$, which acts on the subspace $\HH_{S_n}$
 and is defined by
\begin{align}\label{}
 W_S := T_n P_{S_n}= P_{S_n} \frac{1}{E_n -H^{0}}  {H^I} P_{S_n}.
\end{align}
 Let us use $|\nu\ra$ to indicate the (right) eigenvectors of $W_S$, with eigenvalues $\nu$,
 $W_S |\nu \ra = \nu |\nu\ra$.
 Suppose that $\nu_{\rm max}$ corresponds to the largest absolute value of $\nu $.
 In a generic case, the expansion of $|\psi^{(0)}\ra$ on the basis of $\{ |\nu\ra \}$
 has a nonzero component on $|\nu_{\rm max}\ra$ and, in this case, one has
\begin{align}\label{Ak-eta-k}
 A_k \sim  |\nu_{\rm max}|^{2k} \quad \text{for sufficiently large $k$}.
\end{align}
 Then, the condition (\ref{conv}) is equivalent to
 \begin{align}\label{conv-nu<1}
 \nu_{\rm max} < 1.
\end{align}
\label{footnote-W}}
\begin{align}\label{conv-E0s}
 \lim_{k \to \infty} A_k =0 \quad \forall |\psi^{(0)}\ra = |E^0_{s_0}\ra \in S_n.
\end{align}



\subsection{Perturbative and Nonperturbative parts of EFs}\label{sect-SPT-NPT}

 In a generic analysis, one may intend to consider a set of $S_n$ as large as possible,
 which implies a large size of $|n_P\ra$
 that is expanded in a convergent perturbation expansion [Eq.\eqref{nP-expan-infty}].
 The smallest set $\ov S_n$, for which the condition in Eq.\eqref{conv-E0s} is satisfied,
 was called the {NPT region} of the state $|n\ra$ previously \cite{pre00-GBW,pre02-GBW},
 with the corresponding set $S_n$ as the {PT region} of $|n\ra$.
 In certain limits, it was found that the NPT regions of EFs correspond to
 classically and energetically allowed regions,
 particularly, in the semiclassical limit on the action basis \cite{JPA19-NPT-C},
 and in a spatial discretization limit in the configuration space \cite{CPL04,CPL05}.

 However, for a finite Hilbert space,
 practical computation of an NPT region defined in the above way is usually a hard task,
 both analytically and numerically.
 In fact, from the viewpoint of application, computation of such an NPT region
 is not always necessary.

 Hence, for the practical purpose, one may balance between two requirements on the
 set $\ov S_n$: smallness and simplicity.
 Here, simplicity means that those labels $r$ of $|E_r^0\rangle \in \ov S_n$
 are confined by a simple rule, which facilitates numerical and/or analytical
 computation of the quantity $A_k$ used in Eq.\eqref{conv-E0s}.
 That is, one may impose some confinement rule, which is usually model-dependent
 and motivation-dependent.
 The smallest set $\ov S_n$ that satisfies such a confinement rule
 is to be called a \textit{confined NPT} region, with the related $S_n$ as the confined PT region.
 For brevity,  the word ``confined'' will be usually omitted  in discussions to be given below,
 which has already been done when presenting the main results in Sec.\ref{sect-main-results},
 unless there is a need of emphasizing the confinement.

 For the problem studied here, with respect to the unperturbed energy basis,
 we are to consider the following simple form of $\ov S_n$ (as a confinement), i.e.,
\begin{gather}\label{ovSn-r1r2}
 \ov S_n = \{|E_r^0\ra : r\in [r_1, r_2]\}.
\end{gather}
 The energy width of $\ov S_n$ is to be indicated as  $\Delta_{12}$,
\begin{align}\label{Delta-12}
 \Delta_{12} := E_{r2}^0 - E_{r1}^0.
\end{align}
 From the expression of $T_n$ in Eq.\eqref{T-alpha},
 it is not difficult to see that $E_n$ should usually lie between $E_{r_1}^0$ and $E_{r_2}^0$.
 Moreover, in the model specified in Sec.\ref{sect-model},
 due to the specific form of the matrix of the interaction Hamiltonian,
 we assume without giving a proof that the region $\ov S_n$ is approximately centered at $E_n$, i.e.,
\begin{gather}\label{En=1/2-de}
 \frac 12 \left( E^0_{r_2} + E^0_{r_1} \right) \simeq  E_n .
\end{gather}

 Thus, the NPT region $\ov S_n^{\rm NPT}$ corresponds to the smallest set $\ov S_n$,
 under Eqs.(\ref{conv-E0s}) and (\ref{ovSn-r1r2}),
 and this implies that
\begin{align}\label{wNPT-Delta12}
 w_{\rm NPT} = \min(\Delta_{12}).
\end{align}
 Accordingly, the PT region $S_n^{\rm PT}$ is determined.
 The NPT and PT parts of $|n\ra$, denoted as $|n \ra_{\rm NPT}$ and $|n \ra_{\rm PT}$, respectively,
 are written as
\footnote{ One remark: The semiperturbative theory supplies an approach to the
 PT parts of EFs, as well as division between the PT and NPT parts,
 while, the NPT parts need to be dealt with by some other method.
}
\begin{subequations}\label{|n>PT-NPT}
\begin{align}\label{}
 & |n \ra_{\rm NPT} := Q_{\ov S_n^{\rm NPT}} |n\ra,
\\ & |n \ra_{\rm PT} := P_{S_n^{\rm PT}} |n\ra.
\end{align}
\end{subequations}

 It would be useful to give a brief discussion on some properties of the NPT region $\ov S_n^{\rm NPT}$.
 Under sufficiently weak interactions, for which the ordinary perturbation theory works,
 $\ov S_n^{\rm NPT}$ usually consists of only one unperturbed state
 $|E_{r_0}^0\ra$ whose energy is the closest to $E_n$.
 In this case $r_1=r_2=r_0$.
 Beyond the very weak coupling regime, the size of $\ov S_n^{\rm NPT}$ usually enlarges
 with increasing perturbation strength.

\subsection{GBWPE with finite error}\label{sect-ST-error}

 For the purpose of applying the above discussed semiperturbative theory
 to the main problem to be studied in this paper,
 we need to generalize it to a form that tolerates a given finite error,
 which is denoted by $\epsilon_f$ with ``$f$'' referring to finite.
 This error $\epsilon_f$, though finite, can be very small;
 in fact, in principle, it can be arbitrarily small for a sufficiently large $d_\E$.

 The key observation is that Eq.(\ref{nP-expan}) gives a finite perturbation-expansion
 approximation to $|n_{P}\rangle$,
 if $\| \left( T_n \right)^k |n_P  \rangle \|$ is sufficiently small.
 More exactly, if $A_{k}$ is bounded by $\epsilon_f$, i.e.,
\begin{align}\label{conv-ze-0}
  \sup_{|\psi^{(0)}\ra \in \HH_{S_n}} A_{k} \le \epsilon_f ,
\end{align}
 then,  $|n_P\ra$ has the following approximate expansion,
\begin{align}\label{nP-expan-eps}
 |n_{P}\ra \simeq \sum_{l=1}^{{k}-1} \left( T_n \right)^{l} |n_Q  \rangle,
\end{align}
 for which the norm error is bounded by $\epsilon_f$,
\begin{align}\label{nP-expan-error}
 \left \| |n_{P}\ra - \sum_{l=1}^{k-1} \left( T_n \right)^{l} |n_Q  \rangle \right \|^2
 = \left \|  T_n^{k}  |n_P  \rangle \right \|^2  \le  \epsilon_f.
\end{align}
 The rhs of Eq.(\ref{nP-expan-eps}) gives a finite GBWPE.
 Note that, in the limit of $\epsilon_f \to 0$, Eq.(\ref{conv-ze-0})
 may hold only in the limit of $k \to \infty$, which reduces to Eq.(\ref{conv}).

 It is straightforward to generalize most of the discussions given in the previous
 two sections to the case of finite $\epsilon_f$.
 In particular, the two concepts of NPT and PT parts of EFs are generalizable in a direct way.
 And, corresponding to the GBWPE convergence condition in Eq.(\ref{conv-E0s})
 for the case of $\epsilon_f=0$,
 at a finite $\epsilon_f$, the convergence condition of Eq.(\ref{conv-ze-0}) is simplified to the following one,
\begin{align}\label{conv-ze-0-finite}
  \sup_{|E^0_r\ra \in \HH_{S_n}} A_{k} \lesssim \epsilon_f,
\end{align}
 where ``$\le$'' is loosened to ``$\lesssim$''.
 This implies that the NPT region may be estimated by the following relation,
\begin{align}\label{conv-finite-npt}
  \sup_{|E^0_r\ra \in \HH_{S_n}} A_{k} \approx \epsilon_f \quad \text{for NPT region.}
\end{align}

\section{An approximate expression  of $A_k$ }\label{sect-Ak}

 In this section, we derive an expression for the weight of the $k$-th order term,
 namely $A_k$ in Eq.(\ref{Ak}) for the semiperturbative theory.
 The zeroth-order term $|\psi^{(0)} \ra$ is taken
 as a basis state, namely, $ |\psi^{(0)} \ra = |E^0_{s_0}\ra \in S_n$.
 This section is a crucial step towards the two main results discussed previously.

 Below, we are to consider only those values of $k$ for which $k  < k_M$,
 where $k_M$ indicates a big number satisfying $k_M \ll d_\E$.
 The reason is, as will be explained later,
 that analytical behaviors of the weight $A_k$ at $k \ll d_\E$
 are quite different from those at $k \gg d_\E$ and analysis of the former is simpler than the latter.
 In fact, due to the largeness of $d_\E$,
 for an arbitrarily chosen and fixed value of $\epsilon_f$, which may be very small but not zero,
 it is always possible to adjust the size of the (error-tolerating) NPT region to guarantee validity of
 Eq.(\ref{conv-ze-0-finite}) at sufficiently large $k<k_M$.
 Then, the PT part of the EF has the approximate expansion in Eq.(\ref{nP-expan-eps}).

 Specifically, a formal expression is given for $A_k$ in Sec.\ref{sect-loop-Ak}.
 In Sec.\ref{sect-scaling-analysis}, a scaling analysis is given,
 which simplifies the expression of $A_k$.
 The expression is further simplified in Sec.\ref{sect-Ak-expression}.
 At last, a concise expression of $A_k$ is derived in Sec.\ref{sect-Ak-concise},
 which is useful for later discussions.

\subsection{Loop expression of $A_k$}\label{sect-loop-Ak}

 In this section, a loop expression of the quantity $A_k$ is given.
 Let us consider a component $\psi_r^{(k)}$, which is the projection of the $k$-th order term $|\psi^{(k)}\ra$
 [see Eq.(\ref{psi-(k)})] on a basis state $|E^0_{r}\ra$, i.e.,
\begin{align}\label{psi-r-(k)}
 &\psi_r^{(k)} = \la E_r^0|\psi^{(k)}\ra.
\end{align}
 Clearly, $\psi^{(0)}_{r}= \delta_{rs_0}$ for the zeroth order term.
 According to Eq.(\ref{Ak}),
 \begin{align}
 A_k=&   {\sum_r}'  |\psi_r^{(k)}|^2.
  \label{Ak-F}
\end{align}
 Note that the prime means a summation over the set $S_n$, as defined in Eq.(\ref{prime-sum-r}).
 Making use of Eqs.\eqref{T-alpha} and \eqref{psi-(k)},
 one gets the following iteration  expression,
\begin{gather}
 \psi_r^{(k)}  =  {\sum_{s}}' \frac{ H^{I}_{rs}}{E_n  - E_r^0}   \psi_s^{(k-1)}, \label{Recursion}
\end{gather}
 from the $(k-1)$-th order components to the $k$-th order components.

 We use $s_l$ to indicate the basis labels for components of the $l$-th term $|\psi^{(l)}\ra$,
 namely, $\psi^{(l)}_{s_l} = \la E_{s_l}^0|\psi^{(l)}\ra$ with $l=0, \ldots, k$.
 In this notation, we set $s_k =r$.
 Repeatedly using the relation in Eq.~\eqref{Recursion}  down to the zeroth order,
 one finds that $|\psi_r^{(k)}|^2$ is written as follows,
\begin{align}\label{}\notag
  |\psi_r^{(k)}|^2  =&{\sum_{\substack{s_1, \dots, s_{k-1},\\ t_1, \dots, t_{k-1}}}}'
   \frac{  H^{I}_{s_0s_1}}{(E_n -E^0_{s_1})}
  \frac{ H^{I}_{s_1s_2}}{(E_n -E^0_{s_2})}
  \ldots  \frac{ H^{I}_{s_{k-1}s_k}}{(E_n -E^0_{s_k})}
  \\  \times &  \frac{ H^{I}_{t_k t_{k-1}}}{(E_n -E^0_{t_k})}
 \frac{ H^{I}_{t_{k-1} t_{k-2}}}{(E_n -E^0_{t_{k-1}})}
 \ldots \frac{ H^{I}_{t_1 t_0}}{(E_n -E^0_{t_1})}.
  \label{Kk-ss'r-2}
\end{align}
 where $t_0 = s_0$ and $t_k=s_k = r$.

 It proves convenient to introduce indices $s_l$ with $l>k$, defined as
\begin{align}\label{}
 s_l \equiv t_{2k-l} \quad \text{for $k \le l \le 2k$}.
\end{align}
 Since ${s_{2k}} = t_0 = {s_0}$,
 the $s$-indices form a \textit{loop} with the subscript running from $0$ to $2k$,
 which will be indicated as $\kappa$, i.e.,
\begin{align}\notag
 \kappa :  \  {s_0} \to  {s_1} \to  \dots
 &  {s_{k}}  \to   \dots \to  {s_{2k-1}} \to  {s_{2k}}, \quad
 \\  & \text{ with ${s_{2k}} \equiv {s_0}$ and $s_k = r$}. \label{L12}
\end{align}
 Within such a loop, the index $s_l$ is to be called the $l$-th \textit{point}
 and the sequence of ``$s_{l} \to s_{l+1}$'' the $l$-th \textit{step}.
 Then, Eq.~\eqref{Kk-ss'r-2} is written in a concise form,
\begin{align}
  |\psi_r^{(k)}|^2  =  &{\sum_{\kappa}}'  P(\kappa ),
  \label{Fkr}
\end{align}
 where  $P(\kappa )$ is the contribution of the loop $\kappa$, defined as
\begin{align}
 & P(\kappa ) := \prod_{l=0}^{2k-1} H^{I}_{s_l, s_{l+1}} G_l,  \label{Pd}
 \\ & G_l := \left\{                    \begin{array}{ll} \displaystyle
                      \frac{1}{(E_n -E^0_{s_{l+1}})} & \hbox{for $l<k$,} \\
             \displaystyle         \frac{1}{(E_n -E^0_{s_{l}})} & \hbox{for $l \ge k$.}
                   \end{array}                  \right. \label{Gl}
\end{align}
 Here, $\sum_\kappa'$ represents a summation over all the indices in the loop $\kappa$,
 but not including $s_0= s_{2k}$ and $s_k=r$,
 with the prime as a direct generalization of that introduced in Eq.~\eqref{prime-sum-r};
 more exactly,
\begin{align}\label{sum-kappa}
 {\sum_\kappa}'
 =  {\sum_{s_1}}' \cdots {\sum_{s_{k-1}}}' {\sum_{s_{k+1}}}'  \cdots {\sum_{s_{2k-1}}}'.
\end{align}
 Whenever a summation over $r$ or $s_0$ is needed, it will be written in an explicit way.

 A classification of the loops proves useful,
 which is related to a complex-conjugate relationship between elements of the operator $H^{IA}$
 in the interaction Hamiltonian.
 For this purpose, we introduce a concept,
 which will be referred to as  \textit{environmentally complex-conjugated pair} for steps in a loop,
 in short, \textit{$\E$cc pair}.
 More precisely, two steps within one loop, $s_{l} \to s_{l+1}$ and $s_{l'} \to s_{l'+1}$,
 are said to form an \textit{$\E$cc pair},
 if
\begin{align}\label{il-il'}
  i_{l} = i_{l'+1}   \quad \& \quad    i_{{l}+1} =  i_{l'}.
\end{align}
 Clearly, Eq.(\ref{il-il'}) guarantees that $H^{IA}_{i_{l'} i_{l'+1}} = ( H^{IA}_{i_{l} i_{{l}+1}})^*$
 for an $\E$cc pair,
 such that the environmental part of the contribution of these two steps
 to the rhs of Eq.(\ref{Fkr}) includes $|H^{IA}_{i_{l} i_{l+1}}|^2$.
 A loop is said to be a \textit{fully $\E$cc-paired loop}, if each step
 in it belongs to one and only one $\E$cc pair.

 With respect to $\E$cc pairs, loops are classified into three classes:
 \begin{enumerate}
     \item [(i)] Class I,  containing \textit{fully $\E$cc-paired loops}.
     \item [(ii)] Class II, containing \textit{loops without any $\E$cc pair}.
     \item [(iii)] Class III, containing loops that belong to neither of the above two classes,
 which are referred to as  \textit{partially $\E$cc-paired loops}.
      \end{enumerate}
 Accordingly, the summation on the rhs of Eq.~\eqref{Fkr} is divided into three parts,
\begin{align}\label{F123}
 |\psi_r^{(k)}|^2 =  F^{(k)}_{r1} + F^{(k)}_{r2}  + F^{(k)}_{r3},
\end{align}
 where $F^{(k)}_{r1}$ represents the sum of $P(\kappa )$ over loops in Class I,
 $F^{(k)}_{r2}$ for Class II, and $F^{(k)}_{r3}$ for Class III; that is,
\begin{align}\label{Fk1}
 F^{(k)}_{r1/2/3} = {\sum}'_{\kappa \in \text{Class I/II/III}} \ \ P(\kappa).
\end{align}

\subsection{Fully $\E$cc-paird contribution to $A_k$}\label{sect-scaling-analysis}

 In this section, making use of AR6 [Eq.\eqref{AR6}],
 we discuss scaling behaviors of the three quantities $F^{(k)}_{r1,2,3}$
 with respect to the environmental dimension $d_\E$.
 The result is that $F^{(k)}_{r1}$ dominates in the final contribution and
\begin{align}\label{psir-sim-Fkr1}
 A_k \simeq    {\sum_r}'  F^{(k)}_{r1}
 = {\sum_r}' {\displaystyle \sum_{\kappa \in \text{Class I}}}' \ \ P(\kappa).
\end{align}
 Below, we sketch main ideas that are used for getting Eq.(\ref{psir-sim-Fkr1})
 (See Appendix \ref{app-Fr123-scale} for detailed discussions.)

 In a scaling analysis, due to the largeness of $d_\E$,
 one may neglect the dimension $d_S$ of the small central system $S$,
 in other words, neglect summation over the system's index $\alpha$.
 Moreover, from Eq.(\ref{wNPT-Delta12}) and the assumption AR4 of Eq.(\ref{Lam-LB}),
 one sees that $(\Delta_{12} )^{-1} \ll  \rho^\E_{\rm dos}$, the latter of which scales as $d_\E$
 This implies that, in this analysis, the term $G_l$ in $P(\kappa )$
 [the rhs of Eq.(\ref{Pd})] may also be neglected.

 Firstly, for $F^{(k)}_{r1}$, all the elements of $H^{IA}$ appear in the absolute value form,
 as $|H^{IA}_{i_{l} i_{l+1}}|^2$,
 and there are $(k-1)$ independent $i_l$-indices over which summation is taken.
 Moreover,  due to AR3 of $\lambda < \lambda_{\rm UB}$,
 for each value of $l$ the number of $i_l$ is of the scale of $d_\E$.
 Making use of Eq.(\ref{sigmaIA-estim}) and the fact that the environmental
 density of states is proportional to $d_\E$, one finds the following scaling behavior of
 the matrix elements $H^{IA}_{i_{l} i_{l+1}}$,
\begin{align}\label{HIA-ij-dE-1/2}
 H^{IA}_{i_{l} i_{l+1}} \sim d_\E^{-1/2}.
\end{align}
 Then, after simple derivation, one finds that
\begin{align}\label{Fr1-scale}
 F^{(k)}_{r1} \sim  \frac{1}{d_\E}.
\end{align}

 Secondly, for $F^{(k)}_{r2}$, we consider values of $k$ less than $k_M$ $(k< k_M)$,
 the reason of which is to be explained below.
 Let us consider a set $\gamma$, which consists of all the labels $i_l$ of $s_l \in \kappa$,
 and its subset $\gamma_{\rm sub1} = \{ i_1, \ldots, i_{k-1},i_{k+1}, \ldots, i_{2k-1} \}$,
 under the requirement in Eq.(\ref{gamma-require}).
 Clearly, all the variable $i$-labels in $\kappa$ are contained in $\gamma_{\rm sub1}$.
 Note that, for $k < k_M$, Eq.(\ref{gamma-require}) is valid for steps of most of the loops
 $\kappa$ in Class II
 and, hence, in a scaling analysis, the summation over $\kappa \in \text{Class II}$
 may be replaced by the summation over $\gamma_{\rm sub1}$.
 This implies that $ F^{(k)}_{r2} \sim Q_{\gamma_{\rm sub1}}$, then, according to AR6 [Eq.(\ref{AR6})],
\begin{align}\label{F2-Q1Q2}
 F^{(k)}_{r2} \sim \sqrt{\sum_{i_l \in \gamma_{\rm sub1}} 1}.
\end{align}
 Within the square root on the rhs of Eq.(\ref{F2-Q1Q2}), the summation over
 the $2(k-1)$ $i$-type labels gives a contribution scaling as $d_{\E}^{2(k-1)}$.
 Finally, noting Eq.(\ref{HIA-ij-dE-1/2}), one gets that
 \footnote{\label{novalid_k>kM}
 For $k \gg d_\E$, Eq.(\ref{gamma-require}) is not guaranteed for most of the loops
 and, hence, one may not use AR6 to predict the scaling behavior in Eq.(\ref{Fr2-scale}).
 For this reason, relations to be derived below are not necessarily valid for $A_k$ of $k \gg d_\E$.}
\begin{align}\label{Fr2-scale}
 F^{(k)}_{r2} \sim \frac{1}{d_\E} \quad \text{for $k < k_M$}.
\end{align}
 Although $F^{(k)}_{r1}$ and $F^{(k)}_{r2}$ show similar scaling behaviors with respect to $d_{\E }$,
 since $F^{(k)}_{r2}$ have random signs (in the sense of Eq.(\ref{AR6}) in AR6),
 their contribution to $A_k$ is negligible
 compared with that of $F^{(k)}_{r1}$.

 Thirdly, the quantity $F^{(k)}_{r3}$ may be analyzed by a method of combining
 discussions given above for $F^{(k)}_{r1}$ and $F^{(k)}_{r2}$, leading to
 the same scaling behavior, $F^{(k)}_{r3} \sim d_\E^{-1}$.
 This quantity also has random signs and, as a consequence, its contribution to $A_k$ is
 also negligible.
 Finally, one gets Eq.(\ref{psir-sim-Fkr1}).

\subsection{Ultimate-scheme expression of  $A_k$}\label{sect-Ak-expression}

 In this section, based on the approximate expression of $A_k$ of Eq.(\ref{psir-sim-Fkr1})
 in which all steps are $\E$cc paired,
 we derive an approximate expression of $A_k$
 which is given by an ultimate $\E$cc-pairing scheme.

 To construct loops in Class I  under the pairing condition in Eq.(\ref{il-il'}),
 various schemes of $\E$cc-pairings are available.
 For example, the $0$-th step may be paired with the first step in one scheme,
 while, it may be paired with the second step in another scheme.
 Clearly, the main contribution to $F^{(k)}_{r1}$ comes from
 those schemes, which possess the maximum number of
 independent $i$-indices over which the summation is taken on the rhs of Eq.(\ref{Fk1}).
 Such schemes are to be called \textit{ultimate schemes}.
 Since $s_k=r$ and $s_{2k}=s_0$ are fixed,
 the above mentioned maximum number for ultimate schemes is equal to $(k-1)$.

 To find the ultimate schemes, let us start from the $0$-th step of $s_0 \to s_1$.
 Suppose that it forms an $\E$cc pair with the $l$-th step of $s_{l} \to s_{l+1}$.
 According to Eq.(\ref{il-il'}), their $i$-indices should be connected by the relations of
 $i_{l} = i_1$ and $i_{l+1} =i_0$.
 In the case that the $l$-th step lies in the first half of the loop,
 the largest number of independent $i$-indices
 should correspond to $l=1$, for which the requirement of $i_{l} = i_1$ is satisfied automatically
 and does not impose any further restriction.
 While, in the case of the $l$-th step belonging to the second half of the loop,
 the largest number corresponds to $l= 2k-1$,
 for which the requirement of $i_{l+1} =i_0$ is satisfied automatically.

 In both cases discussed above, $i_1$ is the only independent index over which summation is performed.
 In the first case with $i_2 =i_0$, $P(\kappa)$ in Eq.(\ref{Pd}) contains the following term,
\begin{align}\label{G-prod-1}
 & \frac{1}{E_n-e^\E_{i_0}-e^S_{\alpha_{2}}}
  \sum_{i_1}\frac{1}{E_n-e^\E_{i_1}-e^S_{\alpha_{1}}};
\end{align}
 while, in the second case with $i_{2k-1} =i_1$, it contains
\begin{align}\label{G-prod-2}
 & \sum_{i_1}\frac{1}{(E_n-e^\E_{i_1}-e^S_{\alpha_{1}})(E_n-e^\E_{i_1}-e^S_{\alpha_{2k-1}})}.
\end{align}
 The point lies in that the denominator  in (\ref{G-prod-1}) contains only one $e^\E_{i_1}$,
 while, the denominator  in (\ref{G-prod-2}) contains two $e^\E_{i_1}$.
 Note that $E_n$ lies around the center of the region $\ov S_n$ [Eq.(\ref{En=1/2-de})],
 which implies that the values of $E^0_{s_1}$, as given by
 $E^0_{s_1} = e^\E_{i_1}+e^S_{\alpha_{1}}$, may lie on both sides of $E_n$.
 As a consequence, the value of $|\sum_{i_1}' P(\kappa)|$
 in the first case should be typically much smaller than that of the second.
 Hence, contribution from the first case is negligible, when computing $A_k$ by  Eq.(\ref{psir-sim-Fkr1}).
 \footnote{
 As discussed in Sec.\ref{sect-scaling-analysis} and Appendix \ref{app-Fr123-scale},
 the final contributions of $F_{r2}^{(k)}$ and $F_{r3}^{(k)}$ to $A_k$, which are obtained
 by a summation over $r$, are much smaller than that of $F_{r1}^{(k)}$.
 Here, we note that this is true even at a fixed value of $r$, without the need of taking its summation.
 In fact, most of the terms in $F_{r2}^{(k)}$ and $F_{r3}^{(k)}$ contain
 summations like that in (\ref{G-prod-1}),
 which are much smaller than that in (\ref{G-prod-2}) which is involved in
 the ultimate scheme of $F_{r1}^{(k)}$.
 Hence, $F_{r2}^{(k)}$ and $F_{r3}^{(k)}$ should be
 considerably smaller than $F_{r1}^{(k)}$. \label{F23-small-integrat}
 }

 Following arguments similar to those given above, one finds that
 the main contribution to $A_k$ is given by the case of $i_{2k-2} = i_2$, and so on.
 Finally, one finds that there is only one ultimate scheme of $\E$cc pairing for $A_k$, denoted as $\kappa_c$,
 which is given by
\begin{align}\label{p0:il=i2k-l}
 \kappa_c: \quad  i_{l} = i_{2k-l} \quad \text{for $l=0, \ldots, k-1$}.
\end{align}
 As a result,
\begin{align}\label{Fk1-c}
 A_k \simeq    {\sum_r}' {\sum_{\kappa_c }}' \ \ P(\kappa_c).
\end{align}

 Writing Eq.(\ref{Pd}) explicitly with the ultimate scheme and noting $s_k =r$, one gets that
\begin{align}\notag
   P(\kappa_c) = &
  \frac{|H^{IA}_{i_{k-1} i_{k}}|^2}{(E_n -E^0_{r})^2 }
  \left(  \prod_{l =0}^{2k-1} H^{IS}_{\alpha_{l} \alpha_{l+1}} \right)
\\ & \times \left( \prod_{\substack{l=1 }}^{k-1}
  \frac{|H^{IA}_{i_{l-1} i_{l}}|^2}{(E_n -E^0_{s_{l}}) (E_n -E^0_{s_{2k-l}})} \right).
  \label{Fkr-2}
\end{align}
 When performing the summation over $i$-indices on the rhs of Eq.(\ref{Fk1-c}),
 the squares $|H^{IA}_{i_{l-1} i_{l}}|^2$ on the rhs of Eq.(\ref{Fkr-2}) may be approximated by their variance.
 Then, making use of the assumption of AR5,
 one finds that $P(\kappa_c)$ in this computation is effectively given by
\begin{align}\notag
  & \frac{ \sigma_{IA}^{2k}   \left(  \prod_{l =0}^{2k-1} H^{IS}_{\alpha_{l} \alpha_{l+1}} \right) }
  {(E_n -E^0_{r})^2 \sqrt{\rho^\E_{\rm dos}(e^\E_{i_{0}}) \rho^\E_{\rm dos}(e^\E_{i_{k}})} }
\\ & \times  \prod_{\substack{l=1 }}^{k-1}
  \frac{ 1 }{(E_n -E^0_{s_{l}}) (E_n -E^0_{s_{2k-l}}) \rho^\E_{\rm dos}(e^\E_{i_{l}}) }.
  \label{Fkr-3}
\end{align}
 Finally, substituting Eq.(\ref{Fkr-3}) into Eq.(\ref{Fk1-c}), one gets
 the following approximate expression of $A_k$,
\begin{align}\notag
  A_k =  {\sum_r}' |\psi_r^{(k)}|^2 & \simeq    {\sum_r}'
   \frac{( \rho^\E_{\rm dos}(e^\E_{i_{0}}) \rho^\E_{\rm dos}(e^\E_{i_{k}}) )^{-1/2}}
   {(E_n -E^0_{r})^2}
   \\ & \times \sigma_{IA}^{2k}
   {\sum_{\alpha(\kappa)}} \left(  \prod_{l =0}^{2k-1} H^{IS}_{\alpha_{l} \alpha_{l+1}} \right)  Y^{(k)},
  \label{Fkr2-1-main}
\end{align}
 where ${\sum}_{\alpha(\kappa)}$ is defined in a way similar to the rhs of
 Eq.(\ref{sum-kappa}) and
\begin{align}
 & \label{Y-lambda2}
 Y^{(k)} = \prod_{\substack{l=1 }}^{k-1}
 \left( {\sum_{\substack{i_l}} }' \frac{1 }{(E_n -E^0_{s_{l}})
 (E_n -E^0_{s_{2k-l}}) \rho^\E_{\rm dos}(e^\E_{i_{l}}) } \right),
 \end{align}
 with $s_l = (\alpha_l, i_l)$.
 Note that ${\sum_{s}}'=\sum_{\alpha}{\sum_{i}}'$ for $s=(\alpha,i)$,
 where ${\sum_{i}}'$ is for $ e^\E_i \in (E_n-\frac{e^S_{d_S}+e^S_1}{2}-\frac{b^H_{IA}}{2},
 E^0_{s_1} -e^S_{\alpha})\cup (E^0_{s_2}-e^S_{\alpha},
 E_n-\frac{e^S_{d_S}+e^S_1}{2}+\frac{b^H_{IA}}{2})$; in particular,
 there is no prime for ${\sum}_{\alpha}$.

\subsection{A concise expression of $A_k$} \label{sect-Ak-concise}

 In this section, a concise expression of $A_k$ is derived.
 In fact, under AR4 of  $\lambda > \lambda_{\rm LB}$ which implies that $|G_l | \ll \rho^\E_{\rm dos}$,
 the summation over $i_l$-indices on the rhs of Eq.(\ref{Y-lambda2})
 may be approximated by an integration over the environmental energy, i.e.,
\begin{align}\label{sum-integral}
 {\sum_i}'  \longrightarrow \int ' d e^\E  \rho^\E_{\rm dos}(e^\E) ,
\end{align}
 where the prime over $\int$ means that the integration does not include those
 environmental energy regions, for which the total energy lies between $E_{r1}^0$ and $E_{r2}^0$
 [cf.~Eq.(\ref{ovSn-r1r2})].

Following derivations to be detailed in Appendix \ref{app-compute-Y-lambda}
 we found the following expression of $Y^{(k)}$,
\begin{align}\label{Ykp-expression}
   Y^{(k)} \simeq \left(\frac{4}{\Delta_{12}} \right)^{k-1}
   \ \  \prod_{l=1}^{k-1}  K_{\alpha_l \alpha_{2k-l}},
\end{align}
 where $\Delta_{12}$ is the energy width of $\ov S_n$ [Eq.~\eqref{Delta-12}] and
\begin{widetext}
  \begin{align}
    & \label{abe3} K_{\alpha \beta} := \left\{
      \begin{array}{ll} \displaystyle
         1, & \hbox{for}\  e^S_\alpha = e^S_\beta; \\
         \displaystyle   \frac{\Delta_{12}}{2(e_{\beta}^S-e_{\alpha}^S)}\ln\left(1+2\frac{e^S_{\beta}-e^S_{\alpha}}{\Delta_{12}}\right), & \hbox{for}\  \abs{e^S_\alpha-e^S_\beta}<\Delta_{12}; \\
         \displaystyle   \frac{\Delta_{12}}{2(e_{\beta}^S-e_{\alpha}^S)}
       \ln  \left(1+\frac{\Delta_{12}}{e_{\beta}^S-e_{\alpha}^S-\frac{\Delta_{12}}2} \right), & \hbox{for}\
         \abs{e^S_\alpha-e^S_\beta}>\Delta_{12}.
      \end{array} \right.
   \end{align}
\end{widetext}
 Substituting Eq.(\ref{Ykp-expression}) into Eq.(\ref{Fkr2-1-main}),
 changing the summation over $r$ into integration,
 one gets that
\begin{align}
 A_k =  {\sum_r}' |\psi_r^{(k)}|^2 \simeq & c_A \eta^{k-1} \Z_k,
 \label{Ak=etak-Zk}
\end{align}
 where
\begin{align}\label{eta}
 & \eta := \frac{4  \sigma_{IA}^2}{\Delta_{12}},
 \\ \label{Zk}
 &  \Z_k  :=  {\sum_{\alpha(\kappa)} }
 \prod_{l=0}^{k-1}  H^{IS}_{\alpha_{l} \alpha_{l+1}}
 H^{IS}_{\alpha_{2k-l-1} \alpha_{2k-l}} K_{\alpha_l \alpha_{2k-l}},
 \\ \notag & c_A := \sigma_{IA}^2 {\sum_r}'   \frac{1}{(E_n -E^0_{r})^2}
 \sqrt{\frac{1}{\rho^\E_{\rm dos}(e^\E_{i_{0}}) \rho^\E_{\rm dos}(e^\E_{i_{k}})}}
 \\ & \ \ \ \simeq \sigma_{IA}^2 \int ' d e^\E   \frac{1}{(E_n -E^0_{r})^2}
 \sqrt{\frac{\rho^\E_{\rm dos}(e^\E_{i_{k}})}{\rho^\E_{\rm dos}(e^\E_{i_{0}})}}.
 \label{a}
\end{align}
 with $i_0 \in s_0$ and $i_k \in r$.
 In the above derivation, the fact of $e^S_{\alpha_0} \equiv e^S_{\alpha_{2k}}$ has been used,
 which implies that $K_{\alpha_0 \alpha_{2k}} =1$.

 The quantity $Z_k$ may in fact be written in a concise form.
 For this purpose, let us introduce a label $q$, defined by
 $q = (\alpha, \beta)$, where $\alpha$ and $\beta$ are labels for energy levels of the system $S$.
 With this label, we construct a $d_S^2 \times d_S^2$ matrix $M$,
 whose elements $M_{qq'}$ with $q' = (\alpha', \beta')$ are defined by
\begin{align}\label{}
 M_{qq'} =  H^{IS}_{\alpha \alpha'}
 H^{IS}_{\beta' \beta} K_{\alpha \beta}.
\end{align}
 Then, the product term on the rhs of Eq.(\ref{Zk}) is written as
\begin{align}\label{Mqlql+1}
 M_{q_l q_{l+1}} =
 H^{IS}_{\alpha_{l} \alpha_{l+1}} H^{IS}_{\alpha_{2k-l-1} \alpha_{2k-l}} K_{\alpha_l \alpha_{2k-l}},
\end{align}
 where $q_l = (\alpha_l, \alpha_{2k-l})$
 for $l=0,\ldots, k$.
 Note that $q_0 = (\alpha_0, \alpha_{2k} \equiv \alpha_0)$ and $q_k = (\alpha_k, \alpha_k)$.
 On the rhs of Eq.(\ref{Zk}), the summation over $\alpha(\kappa)$ is by definition
 over $\alpha_1, \ldots, \alpha_{k-1}$ and $\alpha_{k+1}, \ldots, \alpha_{2k-1}$;
 and, hence, it is in fact over $q_1, \cdots, q_{k-1}$.
 Then,  one finds that $ \Z_k   =  \sum_{q_1, \ldots, q_{k-1} }
 M_{q_0 q_1} M_{q_1 q_2} \cdots  M_{q_{k-1} q_{k}}$, more concisely,
\begin{align}
   \Z_k &   = \left( M^{k} \right)_{q_0 q_k}.
\label{Zk-M}
\end{align}

\section{statistical structural properties of EFs}\label{sect-structural-properties-EFs}

 In this section, based on results obtained above,
 we discuss statistical structural properties of EFs,
 particularly, prove the main results discussed in Sec.\ref{sect-main-results}.
 Specifically, the size of NPT region of EFs is discussed in Sec.\ref{sect-NPT-infty-k},
 and PT parts of EFs are discussed in Sec.\ref{sect-EF-infty-dE}.

\subsection{Width of NPT region}\label{sect-NPT-infty-k}

 In this section, we derive an relation that is satisfied by the energy width of NPT region
 and prove the first main result [Eq.(\ref{w-NPT})].

 For large $k$ with $k \sim k_M$, the expression of $Z_k$ in Eq.(\ref{Zk-M}) implies that
\begin{align}\label{Zk-xi}
 Z_k \simeq c(\alpha_0, \alpha_k) \xi^k,
\end{align}
 where $\xi$ is the largest (absolute) eigenvalue of the matrix $M$
 and $c(\alpha_0, \alpha_k)$ is determined by
 positions of the two labels of $q_0$ and $q_k$ in the matrix $M$.
 Note that the $k$-independence of $c$ comes only from the label $\alpha_k$.
 Then, from Eq.(\ref{Ak=etak-Zk}), one gets that
\begin{align}
 A_k\simeq & c_A c(q_0, q_k) \eta^{k-1} \xi^k.
 \label{Ak=etak-Zk-2}
\end{align}

 The NPT width $w_{\rm NPT}$ may be estimated from the relation given in Eq.(\ref{conv-finite-npt}).
 Although $\epsilon_f$ is a small quantity, due to the largeness of $k_M$,
 it is reasonable to assume that $(\epsilon_f)^{1/k} \simeq 1$ for $k \sim k_M$;
 and, similarly, $(c_A c(q_0, q_k))^{1/k} \simeq 1$.
 Then, $w_{\rm NPT}$ should satisfy the following relation
\begin{align}\label{eta-xi=1}
 & {4  \sigma_{IA}^2}  \xi|_{\Delta_{12} =w_{\rm NPT} } \simeq w_{\rm NPT} ,
\end{align}
 where $\xi|_{\Delta_{12} =w_{\rm NPT} }$ means that dependence of $\xi$ on $\Delta_{12}$
 is fixed at $\Delta_{12} = w_{\rm NPT} $.

 Generically, due to the dependence of $\xi$ on $\Delta_{12}$,
 which comes from $K_{\alpha \beta}$ in Eq.(\ref{abe3}),
 it is not an easy task to find an explicit expression of $w_{\rm NPT} $ from Eq.(\ref{eta-xi=1}).
 But, in some cases, dependence of $K_{\alpha \beta}$ on $\Delta_{12}$ is negligible and
 approximate expressions of $w_{\rm NPT}$ are obtainable.

 The first case is for interactions that are relatively weak such that the NPT regions
 are narrow in the sense of $w_{\rm NPT} \ll |e_{\beta}^S-e_{\alpha}^S|$ for
 all $\alpha \ne \beta$.
 In this case, at $\Delta_{12} = w_{\rm NPT} $, according to Eq.~\eqref{abe3},
 $K_{\alpha \beta}  \simeq  \frac{ \Delta_{12}^2}{2(e_{\beta}^S-e_{\alpha}^S)^2} $
 for $\alpha \ne \beta$, much smaller than $K_{\alpha \alpha} =1$.
 As a result, one may neglect the contributions from $\alpha \ne \beta$.
 From Eq.(\ref{Mqlql+1}), one sees that this implies that ${\alpha_l} = {\alpha_{2k-l}}$
 \footnote{That is, in this case, the system's label $\alpha$
 satisfies a scheme similar to the ultimate scheme for the $i$-index in Eq.(\ref{p0:il=i2k-l}).}
 and, as a result,  $ M_{qq'} \simeq  |H^{IS}_{\alpha \alpha'}|^2$.
 Substituting this result into Eq.(\ref{Zk-M}), one finds that
\begin{align}\label{}
 Z_k \simeq  \left[ (m_{\rm w})^k \right]_{\alpha_0 \alpha_k} .
\end{align}
 where $m_{\rm w}$ is a matrix previously defined above Eq.(\ref{w-NPT}).
 Then, following arguments similar to that leading to Eq.(\ref{eta-xi=1}),
 one gets that $ w_{\rm NPT} \simeq 4  \sigma_{IA}^2 \xi_{\rm w} $,
 which is just Eq.(\ref{w-NPT})  with $\xi_{\rm w}$.

 The second case is for interactions that are relatively strong
 such that $w_{\rm NPT} \gg |e_{\beta}^S-e_{\alpha}^S|$ for all $\alpha \ne \beta$.
 In this case,  according to Eq.~\eqref{abe3},
 one has $K_{\alpha \beta}  \simeq    1$ for all $\alpha \ \& \ \beta$
 and, as a result, $ M_{q_l q_{l+1}} \simeq
 H^{IS}_{\alpha_{l} \alpha_{l+1}} H^{IS}_{\alpha_{2k-l-1} \alpha_{2k-l}} $.
 Substituting this result into Eq.(\ref{Zk-M}), one finds that
\begin{align}\label{}
 Z_k \simeq \left[ (H^{IS})^k \right]_{\alpha_0 \alpha_k} \left[ (H^{IS})^k \right]_{\alpha_k \alpha_0}
 = \left|\left[ (H^{IS})^k \right]_{\alpha_0 \alpha_k} \right|^2.
\end{align}
 Then,  one gets Eq.(\ref{w-NPT})  with $\xi_{\rm s}$.

 In the example that all the elements of $H^{IS}$ are equal to $h_S$,
 it is direct to find that $\xi_{\rm w} = d_S h_S^2 $
 and $ w_{\rm NPT}  \simeq  4  \sigma_{IA}^2  d_S h_S^2$ in the first case,
 while,  $\xi_{\rm s} = d_S h_S $
 and $ w_{\rm NPT}  \simeq  4  \sigma_{IA}^2  d_S^2 h_S^2$ in the second case.

\subsection{PT parts of EFs}\label{sect-EF-infty-dE}

 In this section, we discuss PT parts of the EFs and proves the second main result
 discussed in Sec.\ref{sect-main-results}.
 Making use of Eq.~\eqref{nP-expan-eps},  the PT part of the state $|n\ra$ is written as
\begin{align}
  & |n\ra_{\rm PT} \simeq \sum_{k=1}^{k_M}  \left( T_n \right)^{k} |n \ra_{\rm NPT}.
\end{align}
 Expanding the NPT part of the state $|n\ra$ as follows,
\begin{align}\label{}
 | n \ra_{\rm NPT} = \sum_{|E_{s_0}^{0}\rangle\in\overline{S}_{n}^{\text{NPT}}}  C^n_{s_0} |E^0_{s_0} \ra,
\end{align}
 one gets that
\begin{align}
   & |n \ra_{\rm PT}
 \simeq \sum_{k=1}^{k_M} \sum_{|E_{s_0}^{0}\rangle\in\overline{S}_{n}^{\text{NPT}}}  C^n_{s_0} (T_n)^k|E^0_{s_0}\ra.
\end{align}
 Then, the components $C^n_{r} \equiv \langle E_{r}^{0}|n\rangle_{\mathrm{PT}}$ of the PT part are written as
 \begin{equation}\label{Cnr-ErTnkEs0}
 C^n_{r} \simeq \sum_{k=1}^{k_M}\sum_{|E_{s_0}^{0}\rangle
\in\overline{S}_{n}^{\text{NPT}}}C_{s_{0}}^{n}\langle E_{r}^{0} |(T_n)^k|E^0_{s_0}\ra.
 \end{equation}

 Note that the term of $(T_n)^k|E^0_{s_0}\ra$ on the rhs of Eq.(\ref{Cnr-ErTnkEs0})
 is just the vector $|\psi^{(k)}\ra$,
 which has been studied in Sec.\ref{sect-Ak} with the zeroth-order term
 taken as $ |\psi^{(0)} \ra = |E^0_{s_0}\ra \in S_n$.
 As a consequence, $\langle E_{r}^{0} |(T_n)^k|E^0_{s_0}\ra = \psi_r^{(k)}$ [cf.~Eq.(\ref{psi-r-(k)})].

 Further, note that $F_{r2}^{(k)}$ and $F_{r3}^{(k)}$ are in fact
 considerably smaller than $F_{r1}^{(k)}$ (see the footnote \ref{F23-small-integrat}).
 We observe, furthermore, that
 the previous arguments that lead to the relation given in Eq.(\ref{Ak=etak-Zk})
 work even for each fixed label $r$,
 with $c_A$ given by the term under summation in the first equality of Eq.(\ref{a}).
 And, this gives that
 \begin{align}
  \label{Fkr-sim-inf}
|\langle E_{r}^{0}|(T_n)^k|E^0_{s_0}\ra|^{2}
   &  \simeq \frac{1}{(E_n -E^0_{r})^2} D^{(k)}_{rs_0},
\end{align}
 where
\begin{align}
    D^{(k)}_{rs_0} = \left( \rho^\E_{\rm dos}(e^\E_{i_{0}}) \rho^\E_{\rm dos}(e^\E_{i(r)}) \right)^{-1/2}
   \sigma_{IA}^2  \eta_{\rm NPT}^{k-1} \Z_k.
\end{align}
 Here, $\eta_{\rm NPT}$ is the value of $\eta$ taken at $\Delta_{12} = w_{\rm NPT} $, i.e.,
 $ \eta_{\rm NPT}= \frac{4  \sigma_{IA}^2}{w_{\rm NPT}}$.

 Therefore, with its phases denoted by $\theta^{(k)}_{rs_0}$,
 the quantity $\la E^0_r | (T_n)^k|E^0_{s_0}\ra$ has the following expression,
\begin{align}
   \la E^0_r | (T_n)^k|E^0_{s_0}\ra
   &  \simeq \frac{\exp(i\theta^{(k)}_{rs_0})}{(E_n -E^0_{r})}
   \sqrt{ D^{(k)}_{rs_0}}.
  \label{Fkr-sim-inf}
\end{align}
 This gives the expression of the PT part of EF in Eq.(\ref{nr-PT}), namely,
 \begin{align}
    & C^n_r \simeq  \frac{J^n_r}{E_n -E^0_{r}} \quad \text{for $|E^0_r\ra \in S_n^{\rm PT}$, } \tag{\ref{nr-PT}}
  \end{align}
 where
  \begin{align}\label{eq-Jr}
      & J^n_r =  \sum_{s_0 \in  \ov{S}_{\rm NPT}}  \sum_{k=1}^{k_M}  C^n_{s_0} \exp(i\theta^{(k)}_{s_0r})
   \sqrt{ D^{(k)}_{s_0r}}.
  \end{align}
 In particular, the main energy dependence of the EF is given by $(E_n -E^0_{r})^{-1}$.

 Substituting the explicit expression of $T_n$, which is written in terms of
 elements of $H^I$ [as seen from Eq.(\ref{T-alpha})],
 into Eq.(\ref{Fkr-sim-inf}), one may get a formal expression for the phases $\theta^{(k)}_{rs_0}$.
 Since the elements of $H^I$ contain $H^{IA}_{i_l i_{l+1}}$,
 although an expression thus obtained is quite complicated, it is reasonable to expect that,
 due to the phases  $\theta^{(k)}_{rs_0}$ on the rhs of Eq.(\ref{eq-Jr}),
 the quantity $J^n_r$ should typically satisfy AR6 [Eq.(\ref{AR6})].
 As a consequence, when computing local average of $|J_r|^2$ (around $r$),
 offdiagonal contributions may be neglected and
\begin{align}\label{|J|^2=sum}
 & \ov{ |J^n_r|^2}
   \simeq  \frac {\sigma_{IA}^2 B_{\eta \Z}}{ \sqrt{ \rho^\E_{\rm dos}(e^\E_{i(r)}) } }
 \sum_{s_0 \in  \ov{S}_{\rm NPT}} |C^n_{s_0}|^2
 \frac 1{ \sqrt{  \rho^\E_{\rm dos}(e^\E_{i_{0}})} },
\end{align}
 where
\begin{align}\label{}\notag
 B_{\eta \Z} =   \sum_{k=1}^{k_M} \eta^{k-1} \Z_k.
\end{align}

 Now, we compute the summation on the rhs of Eq.(\ref{|J|^2=sum}), written as follows,
\begin{align}\label{}\notag
 & \sum_{s_0 \in  \ov{S}_{\rm NPT}} |C^n_{s_0}|^2  \frac 1{ \sqrt{  \rho^\E_{\rm dos}(e^\E_{i_{0}})} }
  \simeq \sum_{\alpha_0}  \sum_{i_0(NPT)} \frac {|C^n_{s_0}|^2}{ \sqrt{  \rho^\E_{\rm dos}(e^\E_{i_{0}})} }
 \\ & \simeq \sum_{\alpha_0}  \int_{NPT} d e^\E_{i_0} \sqrt{  \rho^\E_{\rm dos}(e^\E_{i_0})} |C^n_{s_0}|^2
 \equiv \sum_{\alpha_0} L_{\alpha_0} ,
\end{align}
 where
\begin{align}\label{Lalpha}
 & L_{\alpha_0} = \int_{E_n - e^S_{\alpha_0}
 - \frac 12 w_{\text{NPT}}}^{E_n - e^S_{\alpha_0} + \frac 12 w_{\text{NPT}}} d e^\E_{i_0}
 \sqrt{ \rho^\E_{\rm dos}(e^\E_{i_0})} |C^n_{s_0}|^2.
 \\ & \simeq \frac{1}{ \sqrt {\rho^\E_{\rm dos}(E_n - e^S_{\alpha_0})}} W^{\rm NPT}_{n \alpha_0}.
\end{align}
 In the above derivation,
 $\rho^\E_{\rm dos}$ was taken as a constant within the NPT window of the width $w_{\text{NPT}}$
 and $W^{\rm NPT}_{n \alpha_0}$ is the NPT region weight of a given label $\alpha_0$, i.e.,
\begin{align}\label{Wn-NPT-weight-alpha}
 & W_{n \alpha} ^{\rm NPT} := \sum_{|E^0_{s} \ra \in  \ov{S}_n^{\rm NPT}} |C^n_s|^2
 \quad \text{for a fixed $\alpha$.}
\end{align}
 Then, one gets that
\begin{align}
 & \ov{ |J^n_r|^2}   \simeq  \frac {\sigma_{IA}^2 B_{\eta \Z}  }
 { \sqrt{ \rho^\E_{\rm dos}(e^\E_{i(r)}) } }
 \sum_{\alpha_0} \frac{W^{\rm NPT}_{n \alpha_0}}{ \sqrt {\rho^\E_{\rm dos}(E_n - e^S_{\alpha_0})}}.
 \label{|Jr|^2-av}
\end{align}

 Making use of the above obtained expression for $J_r$, below, we give an estimate to
 the NPT part weight, $W^{\rm NPT}_{n}$ [Eq.(\ref{Wn-NPT-weight})].
 For this purpose, one may make use of the normalization condition,
\be\label{normalization-PT-NPT}
_{\rm PT}\langle n|n\rangle_{\text{PT}}+W_{n}^{\rm NPT}=1,
\ee
 which implies that
\begin{align}\notag
 &  \sigma_{IA}^2 B_{\eta \Z}  \sum_{\alpha(r)}
 \int_{PT} \frac{d e^\E_{i(r)} \sqrt{ \rho^\E_{\rm dos}(e^\E_{i(r)}) } }{(E_n -E^0_{r})^2}
 \sum_{\alpha_0} \frac{W^{\rm NPT}_{n \alpha_0}}{ \sqrt {\rho^\E_{\rm dos}(E_n - e^S_{\alpha_0})}}
 \\ & \simeq  1-W_{n}^{NPT}.
 \label{int-PT=1-W}
\end{align}
 As an example, in the case that $\rho^\E_{\rm dos}(E_n - e^S_{\alpha_0})
 \simeq \rho^\E_{\rm dos}(E_n )$ for all $\alpha_0$, one gets that
\begin{align}
 &   W_{n}^{\rm NPT} \simeq  \frac 1{1 + \sigma_{IA}^2 B_{\eta \Z}   b_W}
 \label{W=1-W}
\end{align}
 where
\begin{align}
 & b_W =  \sum_{\alpha(r)}
 \int_{PT} \frac{d e^\E_{i(r)}  }{(E_n -E^0_{r})^2}
 \sqrt{ \frac{\rho^\E_{\rm dos}(e^\E_{i(r)})}{ \rho^\E_{\rm dos}(E_n)} }
 \label{T-int}
\end{align}

 It is useful to discuss a simple case, in which
 the environmental density of states may be taken as a constant,
 namely, its average value $\ov\rho^\E_{\rm dos}$.
 Under AR4 which implies that the integration energy domain is much larger than the NPT width,
 one finds that
\begin{align}
 & b_W \simeq  \frac{4 d_S}{w_{\rm NPT}}.
 \label{T-expres}
\end{align}
 In this simple case of constant density of states, $\ov{ |J_r|^2}$ in Eq.(\ref{|Jr|^2-av})
 has the simple expression given below,
\begin{align}
 & \ov{ |J^n_r|^2}   \simeq  \frac {\sigma_{IA}^2 B_{\eta \Z}  }
 { \ov\rho^\E_{\rm dos}} W^{\rm NPT}_{n }.
 \label{|Jr|^2-av-cd}
\end{align}
 Note that the rhs of Eq.(\ref{|Jr|^2-av-cd}) is $r$-independent.
 Then, substituting Eq.(\ref{nr-PT}) into Eq.(\ref{normalization-PT-NPT}),
 and noting AR4, one finds that
\be
 \ov\rho^\E_{\rm dos} \overline{|J^n_r|^{2}}\left(\int_{-\infty}^{-\omega_{\rm NPT}/2}
+\int_{\omega_{\rm NPT}/2}^{\infty}\right)\frac{1}{x^{2}}dx \simeq 1-W_{n}^{\rm NPT},
\ee
 which gives the expression of $\overline{|J^n_r|^{2}}$ in Eq.(\ref{eq-Jr2}).

\section{Some applications}\label{sect-app}

 In this section, we discuss some applications of results obtained above.

\subsection{Thermalization of the central system}\label{sect-thermal-state}

 As a first application,  we discuss the RDM $\rho^S$ of the central system $S$,
 under an eigenstate $|n\ra$ of the total system,
 $\rho^S = {\rm Tr}_\E (|n\ra \la n|)$.
 We first discuss the diagonal elements
 $\rho_{\alpha\alpha}^{S}=\sum_{i}|C_{\alpha i}^{n}|^{2}$, which can be written as
\begin{align}\label{eq-rhod}
 &\rho_{\alpha\alpha}^{S}
 = W_{n\alpha}^{\rm NPT} + W_{n\alpha}^{\rm PT},
\end{align}
 where
\begin{align}\label{}
 & W_{n\alpha}^{\rm NPT} =\sum_{\substack{i\\ |\alpha i\rangle\in\overline{S}_{n}^{\text{NPT}}
} }|C_{\alpha i}^{n}|^{2},
\\ & W_{n\alpha}^{\rm PT} = \sum_{\substack{i\\
|\alpha i\rangle\in S_{n}^{\text{PT}}
}
}|C_{\alpha i}^{n}|^{2}.
\end{align}

 We assume that when computing the above summation for the NPT part,
 due to the largeness of the environmental density of states,
 one may replace $ {|C^n_r|^2}$ by  a smooth function of $E$ indicated as $F_C(E)$,
 at $E=E_r^0 = e^S_\alpha + e^\E_i$.
 Then,
\begin{align*}
 & W_{n\alpha}^{\rm NPT}
 = \int_{E_n - e^S_{\alpha}
 - \frac 12 w_{\text{NPT}}}^{E_n - e^S_{\alpha} + \frac 12 w_{\text{NPT}}} d e^\E_i
 { \rho^\E_{\rm dos}(e^\E_i)} F_C(e^S_\alpha + e^\E_i)
 \\ &  = \int_{E_n  - \frac 12 w_{\text{NPT}}}^{E_n + \frac 12 w_{\text{NPT}}} d E
 { \rho^\E_{\rm dos}(E - e^S_\alpha)} F_C(E)
\end{align*}
 If the density of states behaves exponentially around $E_n$, i.e.,
 $\rho^\E_{\rm dos}(E-e_{\alpha}^{S}) \simeq \rho^\E_{\rm dos}(E)e^{-\beta e_{\alpha}^{S}}$,
 where
\begin{align}\label{beta}
 \beta = \frac{\partial\ln\rho^\E_{\rm dos}(E)}{\partial E}\Big|_{E=E_{n}},
\end{align}
 then,
\begin{align}\label{Wna-NPT-rho}
 & W_{n\alpha}^{\rm NPT}
  \simeq   e^{-\beta e_{\alpha}^{S}}
  \int_{E_n  - \frac 12 w_{\text{NPT}}}^{E_n + \frac 12 w_{\text{NPT}}} d E  \rho^\E_{\rm dos}(E) F_C(E).
\end{align}

 For contribution from the PT region, making use of Eq.~\eqref{nr-PT} and Eq.\eqref{|Jr|^2-av}, one finds that
  \begin{align} \notag
& W_{n\alpha}^{\rm PT}
 \simeq \sum_{\substack{i\\
|\alpha i\rangle\in S_{n}^{\text{PT}}}
}\frac{|J_{\alpha i}|^{2}}{(E_{n}-e_{\alpha}^{S}-e_{i}^{{\cal E}})^{2}}
 \\ & \simeq
 \sum_{\substack{i\\
 |\alpha i\rangle\in S_{n}^{\text{PT}}} }  \frac {\sigma_{IA}^2 B_{\eta \Z}
 \left( \sum_{\alpha_0} \frac{W^{\rm NPT}_{n \alpha_0}}{ \sqrt {\rho^\E_{\rm dos}(E_n - e^S_{\alpha_0})}} \right)
 }
 {(E_{n}-e_{\alpha}^{S}-e_{i}^{{\cal E}})^{2} \sqrt{ \rho^\E_{\rm dos}(e^\E_{i}) }  }.
 \label{eq-rhod1}
  \end{align}
 Replacing the summation over $i$ on the rhs of the above equality into an integral,
 one gets an integration whose integrand has the following
 dependence on $e^S_\alpha$,
\begin{align} \notag
 W_{n\alpha}^{\rm PT}
 \simeq \int_{\rm PT} dE   \sqrt{\rho^\E_{\rm dos}(E-e^S_\alpha)} \times \cdots.
\end{align}
 Clearly, this PT-part contribution could not give rise to a Gibbs factor of $e^{-\beta e^S_\alpha}$,
 though it may contain $e^{-\beta e^S_\alpha/2}$ if the exponential approximation to the
 density of states may hold within a sufficiently large region.

 To summarize the above discussions,
 if the weight of the NPT region dominates, namely
 $W_{n\alpha}^{\rm NPT} \gg W_{n\alpha}^{\rm PT}$,
 which may happen when the interaction is not strong,
 then, the diagonal elements $\rho_{\alpha\alpha}^{S}$
 satisfy the well-known canonical description of Gibbs state, i.e.,
\be
\rho_{\alpha\alpha}^{S} \simeq \frac{e^{-\beta e_{\alpha}^{S}}}{Z},\quad Z=\sum_{\alpha}e^{-\beta e_{\alpha}^{S}}.
\ee
 Otherwise, the behavior of $\rho_{\alpha\alpha}^{S}$ would be more complicated.

 Next, we discuss off-diagonal elements with $\alpha \ne \beta$, which are written as
\begin{align}\label{eq-rhond}
 & \rho^S_{\alpha \beta} = \la \alpha |\rho^S|\beta \ra
 = \sum_i \la \alpha i|n\ra \la n|\beta i \ra = \sum_i C^{n*}_{\beta i} C^n_{\alpha i}.
\end{align}
 We consider a case, in which the interaction is not strong such that
 $|e^S_\alpha - e^S_\beta| > w_{\rm NPT}$.
 In this case, either $|\alpha i\rangle$ or $|\beta i\rangle$ lies in the NPT region and, then,  one gets that
\begin{align} \label{eq-rhooffd}
\rho_{\alpha\beta}^{S} & = \sum_{\substack{i\\
|\alpha i\rangle\in\overline{S}_{n}^{\text{NPT}}
}
}C_{\beta i}^{n*}C_{\alpha i}^{n}+\sum_{\substack{i\\
|\beta i\rangle\in\overline{S}_{n}^{\text{NPT}}
}
}C_{\beta i}^{n*}C_{\alpha i}^{n} \nonumber \\
& \equiv X_\alpha + X_\beta.
\end{align}
 Substituting Eq.~\eqref{nr-PT} into $X_\alpha$, one gets that
\be
X_{\alpha}=\sum_{\substack{i\\
|\alpha i\rangle\in\overline{S}_{n}^{\text{NPT}}
}
}\frac{J_{\beta i}}{(E_{n}-e_{\beta}^{S}-e_{i}^{{\cal E}})}C_{\alpha i}^{n}.
\ee

 For the purpose here, it proves sufficient to give a scaling analysis in $X_\alpha$.
 As discussed previously, $J_{\beta i}$ should have random-type phases in the sense of AR6.
 This implies that
\begin{align}\label{}\notag
  |X_{\alpha}|^{2}\sim & \sum_{\substack{i\\
|\alpha i\rangle\in\overline{S}_{n}^{\text{NPT}}
}
}\frac{|J_{\beta i}|^{2}}{(E_{n}-e_{\beta}^{S}-e_{i}^{{\cal E}})^{2}}|C_{\alpha i}^{n}|^{2}
\\ & \le\frac{\overline{|J_{\beta i}|^{2}}}{(e_{\beta}^{S}-e_{\alpha}^{S})^{2}}\rho_{\alpha\alpha}^{S}.
\end{align}
 From Eq.~\eqref{eq-Jr}, one sees that $\overline{|J_{\beta i}|^{2}}
 \sim (\ov\rho^\E_{\rm dos})^{-1}$.
 Then, one finds that
\be
X_\alpha \sim \frac{1}{\sqrt{\ov\rho^\E_{\rm dos}}}.
\ee
 Similarly $X_\beta \sim \frac{1}{\sqrt{\ov\rho^\E_{\rm dos}}}$.
 Therefore, when the environment is sufficiently large, the RDM has negligible offdiagonal elements.
 In other words, the RDM of the system is diagonal
 on the eigenbasis of the rescaled system Hamiltonian $H^S$.

 To summarise, under a sufficiently large environment and an interaction relatively weak,
 the RDM of the system
 in an eigenstate $|n\rangle$ has a Gibbs form,
\be\label{eq-RDM-final}
\rho^{S}=\text{Tr}_{{\cal E}}(|n\rangle\langle n|)\simeq\frac{1}{Z}e^{-\beta H^{S}}.
\ee
 Thus, the central system is thermalized in one total eigenstate.
 \footnote{ Thermalization in one total eigenstate was first argued in Ref.\cite{Deutch91},
 under certain unproven assumption which may be directly gotten from the second
 main result of this paper.
 }

\subsection{Matrix elements of the central system's operators}

 After deriving the expression of RDM of the system in an eigenstate, in this section,
 we continue to study matrix elements of a generic operator of the central system $S$,
 indicated as  $O^S$.
Within the framework of the ETH ansatz, the matrix element $\langle m|O^{S}|n\rangle$ is written as
\begin{equation}\label{ETH}
    \langle m|O^{S}|n\rangle = \mathcal{O}(e)\,\delta_{mn}
    + \rho_{\text{dos}}(e)^{-1/2} f(e,\omega)\, r_{mn},
\end{equation}
where $e = (E_{m}+E_{n})/2$ and $\omega = E_{m}-E_{n}$.
Here, $\mathcal{O}(e)$ and $f(e,\omega)$ are smooth functions of their respective arguments,
while, $r_{mn}$ are pseudo-Gaussian random numbers with zero mean,
whose variance is equal to $1$ for $n\ne m$ and is usually between $1$ and $2$ for $n=m$.

  We are to study structure of the smooth functions of ${\cal O}(e)$ and of $f(e,\omega)$,
 where we focus on the case of large environment and relatively weak system-environment interaction.
 In fact, the diagonal function follows directly from the
  expression of the RDM given in  Eq.~\eqref{eq-RDM-final}, i.e.,
\begin{align}\label{}
\mathcal{O}(e)
= \overline{\langle n|O^{S}|n\rangle}\big|_{E_{n}\approx e}
\simeq \frac{1}{Z}\,\mathrm{Tr}_{S}\!\left( O^{S} e^{-\beta(e) H^{S}} \right),
\end{align}
where $\beta(e)$ is obtained from Eq.(\ref{beta}) with $E_n$ replaced by $e$.

 Below, we focus on the offdiagonal elements,
\be\label{eq-Hmn}
 \langle n|O^S|m\rangle=\sum_{\alpha, \beta, i} O^S_{\alpha \beta} C_{\alpha i}^{n*}C_{\beta i}^{m},
\ee
 with $m\ne n$.
 For the sake of simplicity in discussion and also for deriving explicit expressions,
 we assume that the environmental density of states may be taken as a constant,
 namely, as $\ov\rho^\E_{\rm dos}$.

 As discussed above, for relatively weak interactions, the main weight of an EF
 is given by its NPT part.
 This implies that, in a study of the main contribution to $\langle n|O^S|m\rangle$
 in Eq.~\eqref{eq-Hmn},
 one may neglect the case in which $|\alpha i\ra$ and $|\beta i\ra$ lie in PT regions of $|E_n\ra$
 and $|E_m\ra$, respectively.
 Thus, we write
\begin{gather}
  \langle n|O^S|m\rangle \simeq \sum_{\alpha, \beta} O^S_{\alpha \beta} (L_1^{\alpha \beta} + L_2^{\alpha \beta} + L_3^{\alpha \beta})
 \end{gather}
  where
 \begin{subequations}\label{}
  \begin{align}\label{}
   & L_1^{\alpha \beta} = \sum_{\substack{i
  \\|\alpha i\rangle\notin\overline{S}_{n}^{\text{NPT}}
   \\ |\beta i\rangle\in\overline{S}_{m}^{\text{NPT}}
  }
  }C_{\alpha i}^{n*}C_{\beta i}^{m} ,
 \\ & L_2^{\alpha \beta} = \sum_{\substack{i\\
  |\alpha i\rangle\in\overline{S}_{n}^{\text{NPT}} \\ |\beta i\rangle\notin\overline{S}_{m}^{\text{NPT}} }
  }C_{\alpha i}^{n*}C_{\beta i}^{m},
  \\ & L_3^{\alpha \beta} = \sum_{\substack{i\\
  |\alpha i\rangle\in\overline{S}_{n}^{\text{NPT}} \\ |\beta i\rangle\in\overline{S}_{m}^{\text{NPT}} }
  }C_{\alpha i}^{n*}C_{\beta i}^{m}.
  \end{align}
  \end{subequations}

 For each fixed pair of $(\alpha, \beta)$, relevant components of the two EFs
 of $C_{\alpha i}^{n}$ and $C_{\beta i}^{m}$ are, in fact, functions of the label $i$
 and the two functions are approximately centered at $(E_m-e^{S}_\beta)$ and $(E_n - e^S_\alpha)$, respectively.
 Therefore, the following two cases needs to be considered separately:
 (i) $\Delta_{mn}^{\beta\alpha} > w_\text{NPT}$ and (ii) $\Delta_{mn}^{\beta\alpha} < w_\text{NPT}$,
 where
\begin{align}
\Delta_{mn}^{\beta\alpha}:=|E_m-e^{S}_\beta-E_n + e^S_\alpha|.
\end{align}

In this paper, we focus only on the case where $\Delta_{mn}^{\beta\alpha} > w_\text{NPT}$. In this regime, it is straightforward to see that $L_3^{\alpha\beta} \simeq 0$.
 Inserting Eq.~\eqref{nr-PT}, one gets that
\begin{subequations}\label{}
\begin{align}
 & L_1^{\alpha \beta} \simeq\sum_{\substack{i\\
|\beta i\rangle\in\overline{S}_{m}^{\text{NPT}}
}
}\frac{(J_{\alpha i}^{n})^{*}C_{\beta i}^{m}}{(E_{n}-e_{\alpha}^{S}-e_{i}^{{\cal E}})}
\\ & L_2^{\alpha \beta} \simeq \sum_{\substack{i\\
|\alpha i\rangle\in\overline{S}_{n}^{\text{NPT}}
}
}\frac{J_{\beta i}^{m}C_{\alpha i}^{n*}}{(E_{m}-e_{\beta}^{S}-e_{i}^{{\cal E}})}.\label{eq-Hmn1}
\end{align}
\end{subequations}
 Without giving further argument, let us assume that signs of the EF components
 within the NPT regions may possess random-type properties, at least in the sense of AR6.
 And, as discussed previously, signs of $J_{\alpha i}$ should be random in the sense of AR6.
 Then, taking local average for the label $i$,
 also over neighboring total states around the energies of $E_n$
 and $E_m$, one finds that
\begin{align} \label{Hnm2-av}
\ov{|\langle n|O^{S}|m\rangle|^{2}} \simeq
 \sum_{\alpha, \beta} |O^S_{\alpha \beta}|^2 \left( \ov{|L_1^{\alpha \beta}|^2}
 + \ov{|L_2^{\alpha \beta}|^2} \right),
\end{align}
 where
 \begin{subequations}\label{}
\begin{align} \label{Lab1^2-sum}
& \ov{|L_1^{\alpha \beta}|^2} \simeq \ov{|J^n_r|^{2} } \sum_{\substack{i\\
|\beta i\rangle\in\overline{S}_{m}^{\text{NPT}} } }
 \frac{  F(E_m,E_r^0) }
{(E_{n}-e_{\alpha}^{S}-e_{i}^{{\cal E}})^{2}},
 \\ & \ov{|L_2^{\alpha \beta}|^2} \simeq \ov{|J^m_r|^{2} } \sum_{\substack{i\\
|\alpha i\rangle\in\overline{S}_{n}^{\text{NPT}}
}
}\frac{ F(E_n ,E_r^0) }{(E_{m}-e_{\beta}^{S}-e_{i}^{{\cal E}})^{2}}.
\end{align}
 \end{subequations}
 Here, $\ov{|J^n_r|^{2} }$ is given in  Eq.\eqref{eq-Jr2} and
 $F(E_m,E_r^0) =\ov{|C_{\alpha i}^{m}|^{2} }$, indicating the envelope function
 of the average shape of EFs.
 Replacing the summation by integration in Eq.(\ref{Lab1^2-sum}), one gets that
\begin{align} \label{Lab1^2-inte}
& \ov{|L_1^{\alpha \beta}|^2} \simeq \ov{|J^n_r|^{2} }
 \int_{E_m  - \frac 12 w_{\text{NPT}}}^{E_m + \frac 12 w_{\text{NPT}}} d E^0
 \frac{ \ov\rho^\E_{\rm dos} F(E_m,E^0) }{(E_{n} - E^0)^2 }.
\end{align}

 To estimate Eq.~\eqref{Lab1^2-inte}, we assume that the envelope
 function $F(E_m,E_r^0)$ is flat within the NPT region,
 indicated as $F(E_m)$.
 Then, with straightforward calculations, one gets
 \begin{gather}\label{Lab1^2-fin-1}
  \ov{|L_1^{\alpha \beta}|^2} \simeq \ov{|J^{n}_r|^2}  \ov\rho^\E_{\rm dos}
\frac{F(E_{m})w_{{\rm NPT}}}{(\Delta_{mn}^{\beta\alpha})^{2}-\frac{w_{{\rm NPT}}^{2}}{4}}.
\end{gather}
To go further, we note that $W_m^{\rm NPT}$ has the following expression,
\begin{align}\label{}
 W_{m\alpha} ^{\rm NPT}  =
 \int_{E_m  - \frac 12 w_{\text{NPT}}}^{E_m + \frac 12 w_{\text{NPT}}} d E^0
 \ov\rho^\E_{\rm dos} F(E_m,E^0),
\end{align}
 which is $\alpha$-independent because $\ov\rho^\E_{\rm dos}$ has been taken as a constant.
 Then, one has
\begin{align}\label{F(E_m)}
  F(E_m) = \frac{ W_{m} ^{\rm NPT}  }{ d_S w_{\text{NPT}}  \ov\rho^\E_{\rm dos}},
\end{align}
 Inserting Eq.(\ref{F(E_m)}) into Eq.(\ref{Lab1^2-fin-1}) and making use of Eq.\eqref{eq-Jr2},
 one gets that
\begin{gather}\label{Lab1^2-fin}
  \ov{|L_1^{\alpha \beta}|^2} \simeq
 \frac{1}{d_S \ov\rho^\E_{\rm dos}}
\frac{ \omega_{\text{NPT}}\left(1-W_{n}^{\rm NPT}\right) W_{m} ^{\rm NPT}  }{(\Delta_{mn}^{\beta\alpha})^{2}-\frac{w_{\rm NPT}^2}{4}}.
\end{gather}

 Similarly, an expression of $\ov{|L_2^{\alpha \beta}|^2}$ can be derived,
 which is given by the rhs of Eq.(\ref{Lab1^2-fin}) with an exchange of $m \leftrightarrow n$.
Then, from Eq.(\ref{Hnm2-av}), in case of $\omega \gg w_{\text{NPT}}$ and $\omega \gg |e^S_\beta - e^S_\alpha|$,  one gets the following dependence of the envelope function
 $f(e,\omega)$ on $\omega = E_m - E_n$,
\begin{gather}\label{feomega-sim}
 |f(e,\omega)|^2 \sim
\frac{ 1 }{(\omega)^2-\frac{w_{\rm NPT}^2}{4}} \sim \frac{1}{\omega^2} .
\end{gather}

We recall that, in the model studied above, all the derivations are done
within the subspace spanned by those unperturbed states $\ket{E^0_r}$ satisfying $|E^0_r - E_0| \le b^H_{IA}$,
with $b^H_{IA}$ as the band width of the interaction Hamiltonian $H_I$ in the original model.
Hence, Eq.~\eqref{feomega-sim} indicates that
\be
|f(e,\omega)|^{2}\sim\frac{1}{\omega^{2}}\ ,\text{for}\ \omega_{\text{NPT}}\ll\omega<b_{IA}^{H}\ .
\ee


\begin{figure}[!t]
 \centering
\includegraphics[width=1\linewidth]{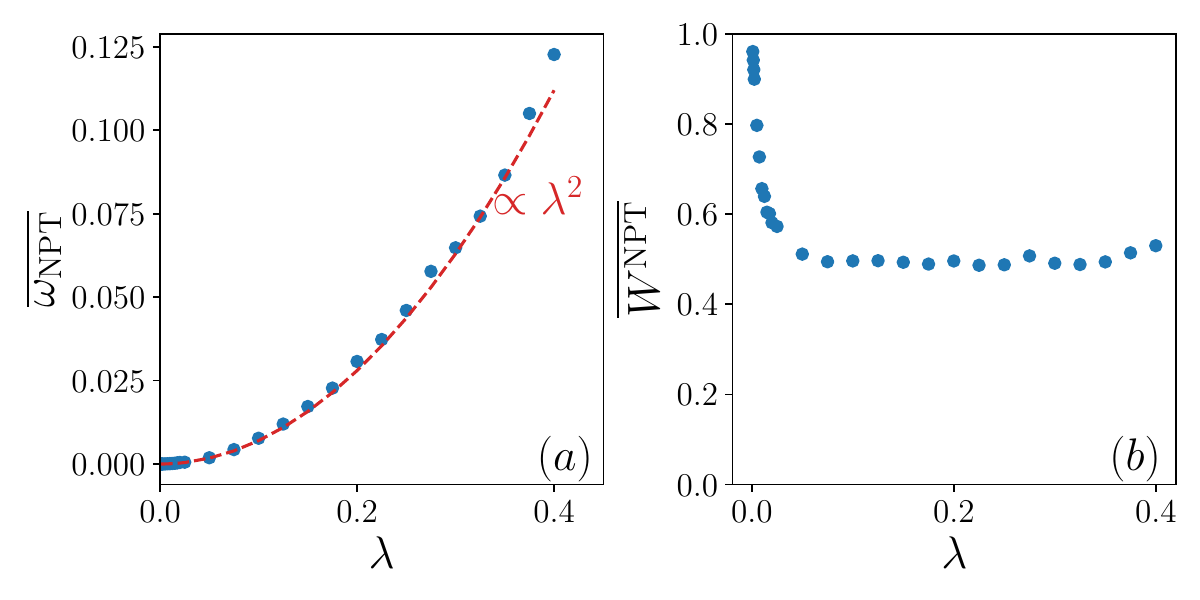}
  \caption{(a) Averaged width of NPT region $\overline{\omega_{\text{NPT}}}$
  and  (b)  averaged weight of the NPT region $\overline{W^{\text{NPT}}}$
  versus coupling strength $\lambda$. The dashed line in (a) indicates predicted the scaling $\propto \lambda^2$ given in Eq.~\eqref{w-NPT}.
The results are obtained by averaging over $10$ eigenstates in the middle of the spectrum, and over $10$ different realizations of $H^{I\mathcal E}$ and $H^{\mathcal E}$.
  }
  \label{Fig1}
\end{figure}

\section{Numerical simulations}\label{sect-numerics}
In this section, we present numerical results to validate our analytical predictions, where we focus on Eqs.~\eqref{w-NPT}, \eqref{nr-PT}, and~\eqref{feomega-sim}.
To this end,  we consider a model,  with a structure similar to that is introduced in Sec.~\ref{framework}. The Hamiltonian reads,
\begin{equation}\label{H_Numerical}
H=\omega_S\sigma_{z}+\lambda\sigma_{x}H^{I{\cal E}}+H^{{\cal E}},
\end{equation}
where $\sigma_{x,z}$ represent the Pauli matrices and $\lambda$ characterizes the interaction strength.  $H^{\cal E}$ and $H^{I{\cal E}}$ are $d_{\cal E} \times d_{\cal E}$ matrices:
 $H^{I\mathcal E}$ is a random matrix sampled from the Gaussian Orthogonal Ensemble (GOE), whose matrix elements have variance $1/(4 d_{\mathcal E})$;
$H^{\mathcal E}$ possesses a constant density of states, with its eigenvalues chosen uniformly from the interval $[-1,1]$. Throughout our numerical simulations we fix $\omega_S = 0.1$ and $d_{\cal E} = 2^{13}$.

First, let us consider Eq.~\eqref{w-NPT}, which predicts that $\omega_{\text{NPT}} \propto \sigma_{IA}^{2}$.  
According to Eq.~\eqref{sigmaIA-estim}, and for the model employed here [see Eq.~\eqref{H_Numerical}], one has $\sigma_{IA}^{2} \propto \lambda^{2}$.  
In Fig.~\ref{Fig1} (a), we compute the width of the NPT region $\omega_{\text{NPT}}$ (see Appendix \ref{sect-ct-npt} for more details), for different coupling strengths $\lambda$, averaging over $10$ eigenstates in the middle of the spectrum and over $10$ different realizations of $H^{\mathcal E}$ and $H^{I\mathcal E}$.  
The scaling $\overline{\omega_{\text{NPT}}} \propto \lambda^{2}$ is clearly observed, suggesting the validity of Eq.~\eqref{sigmaIA-estim}. Moreover, the averaged weight of the NPT of the EFs, denoted by $\overline{W^{\text{NPT}}}$, is shown in Fig.~\ref{Fig1} (b), where 
an approximately constant $\overline{W^{\text{NPT}}}$ is observed at intermediate coupling strength $\lambda \ge 0.1$.

\begin{figure}[!t]
  \centering
  \includegraphics[width=1\linewidth]{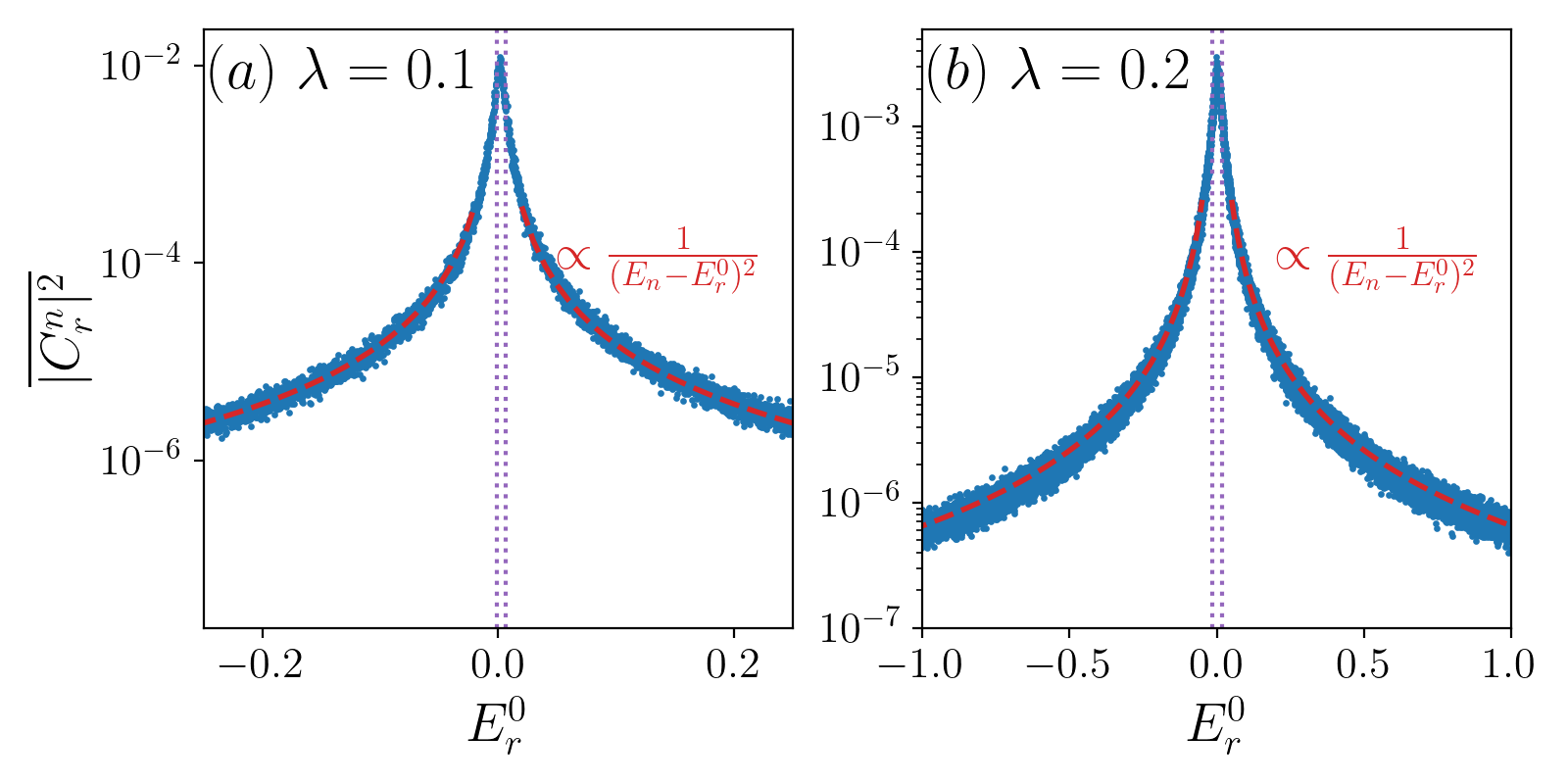}
  \caption{Averaged shape of eigenfunctions $\overline{|C_{n}^{r}|^{2}}$ for (a) $\lambda = 0.1$ and (b) $\lambda = 0.2$.  The dashed line indicate the predicted scaling $\propto(E_{n}-E_{r}^{0})^{-2}$ given in Eq.~\eqref{nr-PT}.
  The results are obtained using a single eigenstate in the middle of spectrum and averaging over $100$ independent realizations of $H^{I\mathcal E}$ and $H^{\mathcal E}$.
}
  \label{Fig2}
\end{figure}

Next, we study Eq.~\eqref{nr-PT}, in which we evaluate the average shape of the eigenfunctions, denoted by $\overline{|C_{n}^{r}|^{2}}$.  
More precisely, we focus on a single eigenstate in the middle of the spectrum and average over $100$ different realizations of the Hamiltonians.  
As shown in Fig.~\ref{Fig2}, a power-law scaling $\overline{|C_{n}^{r}|^{2}} \propto (E_{n} - E_{r}^{0})^{-2}$ is clearly observed, in agreement with the prediction of Eq.~\eqref{nr-PT}.

\begin{figure}[!t]
  \centering
  \includegraphics[width=1\linewidth]{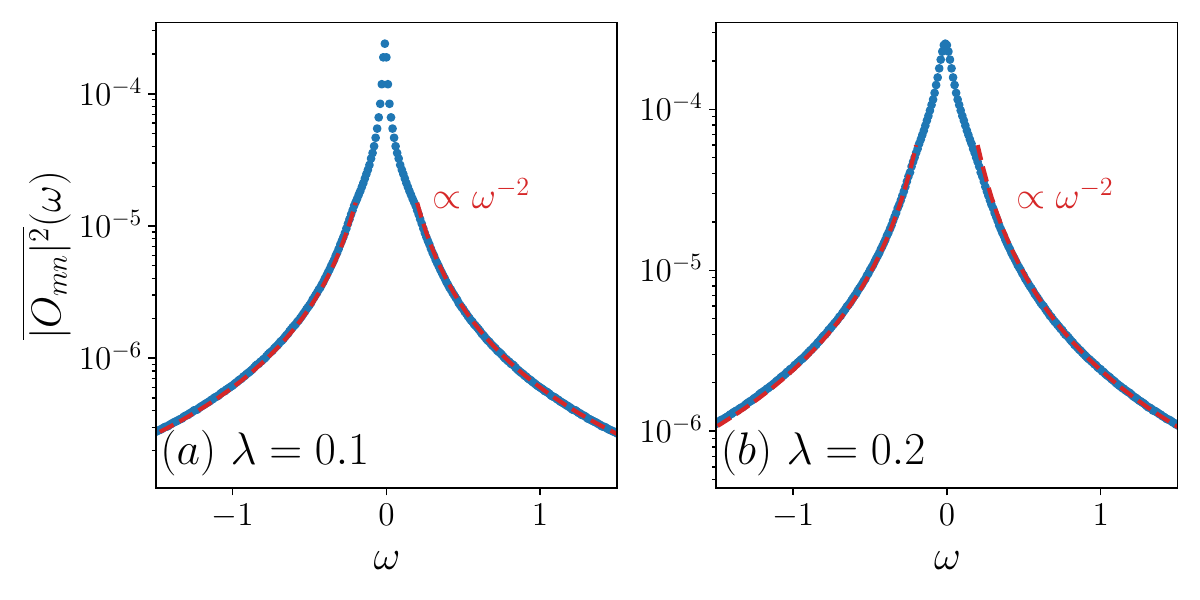}
  \caption{Averaged off-diagonal elements of operator $O=\sigma_z$ for (a) $\lambda = 0.1$ and (b) $\lambda = 0.2$. The dashed line indicates the scaling $\propto \omega^{-2}$ given in Eq.~\eqref{feomega-sim}.}
  \label{Fig3}
\end{figure}

After verifying the two main analytical predictions of our paper, we move on to examine one application of our results, given in Eq.~\eqref{feomega-sim}.
Here we consider the operator $O=\sigma_z$ as an example and study
\begin{equation}
    \overline{|O_{mn}|^{2}}(\omega)=\frac{1}{N_{\omega}}\sum_{\substack{m,n\\
|E_{m}-E_{m}|\approx\omega
}
}|O_{mn}|^{2},
\end{equation}
where the sum runs over the $N_\omega$ matrix elements satisfying
\begin{gather}
|E_m - E_n| \in \left[\,\omega - \frac{\Delta\omega}{2},\ \omega + \frac{\Delta\omega}{2}\,\right], \nonumber \\
\left|\frac{E_m + E_n}{2} - E_0\right| \le \Delta E.
\end{gather}
In our numerical simulations we choose $E_0 = 0$, $\Delta E = 0.1$, and $\Delta\omega = 0.01$, and results are shown in Fig.~\ref{Fig3}.
For large $\omega\gg 0$, the data reveals the asymptotic scaling behavior
$\overline{|O_{mn}|^{2}}(\omega)\propto\omega^{-2}$, in good agreement with Eq.~\eqref{feomega-sim}.

\section{Summary and discussions}\label{sect-conclusion}
\subsection{Summary}

In this work, in a model of a generic type though with certain restrictions specified below,
we have analytically investigated statistical structural properties in the main bodies of the
EFs (energy eigenfunctions) of a composite quantum system, where a central (relatively small) system
is locally coupled to a (huge) chaotic environment.
Here, statistical structural properties of EFs refer to
their average shape and statistical fluctuations,
on the basis which is given by the direct product of the energy eigenbases of the central system and the environment.
Locality of the interaction refers to the environmental part, which implies that the environmental part of
the interaction Hamiltonian has a band structure on the environment's eigenbasis, with a bandwidth $b^{H}_{IA}$.

The studied model fulfills the following main restrictions.
Firstly, the interaction strength is restricted to an intermediate regime,
which is not very weak such that it may induce thermalization and decoherence,
meanwhile not strong such that the main bodies of the EFs are much narrower than
the bandwidth $b^{H}_{IA}$.
In this regime of interaction strength, to study the total EFs within a narrow energy shell,
one may focus on a limited environmental energy region,
within which the environmental part of the interaction Hamiltonian matrix, $H^{IA}_{ij}$,
is full on the eigenbasis.

Secondly, the main property of the environment, which is assumed and related to its chaotic dynamics,
is that the elements $H^{IA}_{ij}$ fluctuate around zero and the sum of their products
scales as the square root of the number of products in the sum.
Although the latter behavior is similar to random numbers,
correlations among $H^{IA}_{ij}$ are not forbidden.

A semi-perturbative theory has been employed in the analytical study of
statistical structural properties of the EFs in the model.
Within this theory, each EF is divided into a PT (perturbative) part and an NPT (nonperturbative) part,
where a major weight of the EF lies in the NPT part when the interaction is not very strong.
An analytical expression has been derived for the PT parts,
which predicts both the average decaying behavior of this part --- a power-law decay ---
and the fluctuation behaviors.
Furthermore, for both relatively weak and relatively strong interactions,
approximate expressions have been derived for the width of the NPT part;
meanwhile, for interaction in between, a relation has been derived which should be satisfied by the width.

Two applications of the above results to two fundamental problems are discussed.
One is a study of the RDM of the central system in one total eigenstate of the composite system,
predicting a condition under which the RDM may have a Gibbs form.
The other lies in the derivation of an expression of
the smooth off-diagonal function for operators of the central system within the framework of ETH,
predicting a power-law decaying behavior of the function.

\subsection{Outlook}

This work can be regarded as a first step toward a final goal,
which is to understand statistical structural properties of the EFs in models
that are not restricted by the above-mentioned upper bound for the interaction strength
and not limited to the main bodies of the EFs.
To achieve this goal, however, is a highly nontrivial task,
particularly due to the banded structure of the interaction Hamiltonian.
To make the problem more tractable, it is useful to divide it into several subproblems:
\begin{enumerate}
  \item[(a)] To investigate the statistical properties within the environmental subspace
  spanned by those unperturbed states that lie within a bandwidth $b^{H}_{IA}$.

  \item[(b)] To extend the analysis to the entire Hilbert space, assuming
  a constant bandwidth $b^{H}_{IA}$, independent of the environmental energy region,
  and taking into account specific contributions from the small central system.

  \item[(c)] To further generalize the problem by allowing $b^{H}_{IA}$ to vary slowly with energy
  and by permitting some of the interaction Hamiltonian elements $H^{IA}_{ij}$
  to vanish within the band.
\end{enumerate}
The work of this paper takes a first step by focusing on subproblem (a).
Extension to the cases of (b) and (c) is an interesting subject for future investigations.

After a clearer understanding of these subproblems has been achieved,
a natural next step would be to employ our framework to study thermalization and decoherence processes
starting from generic nonequilibrium initial states.
It is also of interest to explore situations where the two subsystems are of comparable size,
so that energy or magnetization exchange can be addressed within the same framework.





\acknowledgements

W.-g. Wang would like to thank Ming Yuan for his involvement in an initial stage of this work,
when he was an undergraduate student in USTC. 
 This work was partially supported by the Natural Science Foundation of China under Grant
 Nos.~12175222, 92565306, and 11775210. JW acknowledges support from the Deutsche
Forschungsgemeinschaft (DFG), under Grant No. 531128043.

\appendix

\section{Definition of main-body region of EF}\label{sect-main-body-region}

 In this appendix, we recall a definition of main-body regions of EFs.

 In fact, significant components $C_r^n$ usually occupy a restricted region in the uncoupled
 spectrum, say, in a region of $ E_r^0$ with $r$ between $r_{n}^L$ and $r_{n}^H$.
 We call such a region the \textit{``main-body'' region} of the state $|n\ra$ on the basis of $|E_r^0\ra$
 and denote it by $\Gamma_n^{\rm mb}$,
\begin{align}\label{}
 \Gamma_n^{\rm mb} = [r_{n}^L, r_{n}^H].
\end{align}
 To characterize quantitatively, one may employ a small positive parameter $ \epsilon_{\rm mb}$,
 such that the weight of $|n\ra$ outside this region is smaller than $ \epsilon_{\rm mb}$, i.e.,
 \footnote{The exact value of $ \epsilon_{\rm mb}$ is usually case-dependent.
 That is, it depends on what is needed for the problem at hand; it may be, say, $10\%$, or $1\%$.
 But, the name of ``main body'' implies that $ \epsilon_{\rm mb} $ should not be very small. }
\begin{gather}\label{main-body}
 \sum_{r \in \Gamma_n^{\rm mb}} |\la E_r^0|n\ra |^2 \doteq 1- \epsilon_{\rm mb},
\end{gather}
 where ``$\doteq$'' means that the left-hand side is either equal to the rhs, or is just larger than  the rhs,
 such that it become smaller than the rhs when $[r_{n}^L, r_{n}^H]$ is shrunk
 by letting $r_{n}^L \to r_{n}^L +1$ or $r_{n}^H \to r_{n}^H -1$.

 We use $\D^{\rm mb}_n$ to denote the domain of the energy region occupied by $\Gamma_n^{\rm mb}$, i.e.,
\begin{gather}\label{w-E}
 \D^{\rm mb}_n=   E^0_{r_{n}^H} - E^0_{r_{n}^L}.
\end{gather}
 Apart from the error $\epsilon_{\rm mb}$, the scale of $\D^{\rm mb}_n$ is restricted by
 the energy scope of the interaction, i.e., by $\| H^I \|_{S+A}$, where the subscript ``$S+A$''
 means that the norm is taken within the relevant state space of the two systems of $S$ and $A$.
 More explicitly, one may write
\begin{align}\label{}
 \D^{\rm mb}_n = a \| H^I \|_{S+A},
\end{align}
 where $a$ is some prefactor that depends on $\epsilon_{\rm mb}$.

\section{Proof of Eq.(\ref{sigmaIA-estim})}\label{app-prove-sigmaIA}

 In this section, we give a proof for the estimate to $\sigma_{IA}^2$ given in Eq.(\ref{sigmaIA-estim}).
 In this estimate, we take the environmental density of states as a
 constant, denoted by $\ov \rho^\E_{\rm dos}$.

 According to AR5, $\sigma_{IA}^2$ is written as follows,
\begin{align}\label{}\notag
 & \sigma_{IA}^2 = \frac{1}{M} \sum_i
 \sum_{j: |e^\E_i - e^\E_j| < b^{H}_{IA}} |H^{IA}_{ij}|^2 \sqrt{\rho^\E_{\rm dos}(e^\E_i) \rho^\E_{\rm dos}(e^\E_j) }
 \\ & \simeq  \frac{1}{M}  \ov \rho^\E_{\rm dos} \sum_i
 \sum_{j: |e^\E_i - e^\E_j| < b^{H}_{IA}} |H^{IA}_{ij}|^2,
\end{align}
 where
\begin{align}\label{}
 M =  \sum_i \sum_{j: |e^\E_i - e^\E_j| < b^{H}_{IA}} 1
 \simeq 2 d_A d_{B} \ov \rho^\E_{\rm dos}  b^{H}_{IA}.
\end{align}
 Note that
\begin{align}\label{}
 & \tr_\E(H^{IA})^2 = d_{B}  \tr_A (H^{IA})^2,
\end{align}
 meanwhile,
\begin{align}\label{}
 &  \tr_\E(H^{IA})^2 = \sum_{i,j} |H^{IA}_{ij}|^2.
\end{align}
 Hence, making use of AR2, one gets that
\begin{align}\label{}
   \tr_A (H^{IA})^2 d_{B} = \sum_i \sum_{j: |e^\E_i - e^\E_j| < b^{H}_{IA}} |H^{IA}_{ij}|^2.
\end{align}

 Finally, from the above results, one gets that
\begin{align}\label{}
 & \sigma_{IA}^2
 \simeq  \frac{\ov \rho^\E_{\rm dos}  \tr_A (H^{IA})^2 d_{B}}{2d_A d_{B} \ov \rho^\E_{\rm dos}  b^{H}_{IA}}
  =  \frac{ \tr_A (H^{IA})^2}{2d_A  b^{H}_{IA}},
\end{align}
This proves Eq.(\ref{sigmaIA-estim}).

\section{Proof of Eq.(\ref{psir-sim-Fkr1})}\label{app-Fr123-scale}

 Firstly,  let us discuss the quantity $F^{(k)}_{r1}$, which comes from loops in Class I
 according to its definition in Eq.(\ref{Fk1}).
 By definition, a fully $\E$cc-paired loop with $2k$ steps consists of $k$ independent $\E$cc pairs,
 for each of which the elements of $H^{IA}$ appear in the absolute value form,
 as $|H^{IA}_{i_{l} i_{l+1}}|^2$.
 Since $s_{0}$ and $s_k$ are fixed, the number of independent
 $i_l$-indices in $F^{(k)}_{r1}$  is equal to $(k-1)$.
 Hence, the summation over $\kappa$, which is performed on the rhs of Eq.(\ref{Fk1}) for $F^{(k)}_{r1}$,
 is for these independent $i_l$-indices.

 As indicated by the prime on the rhs of Eq.(\ref{Fk1}) [cf.~Eq.(\ref{prime-sum-r})],
 for each $i$-index, the summation is taken over those states $|i\ra$ for which $|E^0_s\ra \in S_n$.
 Note that, under AR3 of $\lambda < \lambda_{\rm UB}$,
 the set $S_n$ of interest is much larger than $\ov S_n$.
 As a consequence, the summation for each $i$-index is almost over
 the whole space of $\HH_\E$ and gives a result approximately proportional to $d_\E$.
 In the scaling analysis, one may make use of the estimate to $\sigma_{IA}^2$
 given in Eq.(\ref{sigmaIA-estim}),
 which together with the fact of $\rho^\E_{\rm dos} \sim 1/d_\E$ implies that
 $|H^{IA}_{i_{l} i_{l+1}}|^2 \sim d_\E^{-1}$.
 Then, one finds that $F^{(k)}_{r1}$ should scale as
\begin{align}\label{Fr1-scale-app}
 F^{(k)}_{r1} \sim \frac{d^{k-1}_{\E }}{d_\E^k} = \frac{1}{d_\E}.
\end{align}

 Secondly, we discuss the scaling behavior of $F^{(k)}_{r2}$ for loops in Class II,
 which in fact contains most of the terms on the rhs of Eq.~\eqref{Fkr}.
 For this purpose, we impose a restriction to the values of $k$ to be considered;
 that is, $k< k_M$, where $k_M$ is big number satisfying $k_M \ll d_\E$.
 As a merit of this restriction, within most of the $2k$-step loops with $k<k_M$,
 the elements $H^{IA}_{ij}$ are not equal to each other, nor to their complex conjugates.
 And, as a result, such a loop satisfies the requirement in Eq.(\ref{gamma-require}),
 which enables usage of the assumption AR6.

 To be specific,
 let us consider a set $\gamma$, which consists of all the labels $i_l$ of $s_l \in \kappa$,
 and its subset $\gamma_{\rm sub1} = \{ i_1, \ldots, i_{k-1},i_{k+1}, \ldots, i_{2k-1} \}$,
 under the requirement in Eq.(\ref{gamma-require}).
 Clearly, all the variable $i$-labels in $\kappa$ are contained in $\gamma_{\rm sub1}$.
 Since  Eq.(\ref{gamma-require}) is valid for steps of most of the loops $\kappa$ at $k < k_M$,
 in a scaling analysis, the summation over $\kappa \in \text{Class II}$
 may be replaced by the summation over $\gamma_{\rm sub1}$.
 More exactly, substituting Eq.(\ref{Pd}) with $G_l$ omitted into Eqs.(\ref{Fk1}) of $F^{(k)}_{r2}$
 and noting Eq.(\ref{QM}), one finds that $ F^{(k)}_{r2} \sim Q_{\gamma_{\rm sub1}}$.
 Then, according to Eq.(\ref{AR6}) of AR6, one gets that
\begin{align}\label{F2-Q1Q2-app}
 F^{(k)}_{r2} \sim \sqrt{\sum_{i_l \in \gamma_{\rm sub1}} 1}.
\end{align}
 Finally, making use of the above result and taking into account the contribution from
 the elements $H^{IA}_{ij}$, which as discussed above scales as $H^{IA}_{ij} \sim d_\E^{-1/2}$,
 one gets the following scaling behavior of $F^{(k)}_{r2}$,
\begin{align}\label{Fr2-scale-app}
 F^{(k)}_{r2} \sim \frac{1}{d_\E} \quad \text{for $k < k_M$}.
\end{align}

 Although $F^{(k)}_{r1}$ and $F^{(k)}_{r2}$ show similar scaling behaviors with respect to $d_{\E }$,
 there is a big difference between their final contributions to
 the quantity $A_k$, when the summation on the rhs of Eq.~\eqref{Ak-F} is taken.
 And, as a result, the contribution of $F^{(k)}_{r2}$ to $A_k$ should be finally negligible,
 compared with that of $F^{(k)}_{r1}$.

 To see the above mentioned point, let us divide the relevant energy region of $E^0_r$
 into a series of narrow windows, denoted by $\Gamma_a$ of $a=0,1,\ldots$,
 which contain $N_{\Gamma_a}$ levels ($N_{\Gamma_a} \gg 1$)
 and are centered at points $E^0_{\Gamma_a}$, respectively.
 Here, narrowness of $\Gamma_a$ means that $E^0_r \simeq E^0_{\Gamma_a}$
 for all $E^0_r \in \Gamma_a$.
 Note that the number of the narrow windows
 increases polynomially with the particle number $N_\E$ due to polynomial increase of energy,
 while, the environmental density of states increases exponentially with $N_\E$.
 As a result, $N_{\Gamma_a}$ should scale as $d_{\E }$.
 Further, Since the signs of $F^{(k)}_{r1}$ are not random in any sense,
 taking the summation over $r$ of $E^0_r \in \Gamma_a$,
 the contribution of $F^{(k)}_{r1}$ to $A_k$ in Eq.~\eqref{Ak-F} should scale as $N_{\Gamma_a}$.
 While, $F^{(k)}_{r2}$ should possess random signs at least in the sense of AR6.
 This implies that the contribution of $F^{(k)}_{r2}$ should scale as $N_{\Gamma_a}^{1/2}$.
 As a consequence, the contribution of $F^{(k)}_{r2}$ to $A_k$ is finally negligible.

 Thirdly, we discuss the quantity $F^{(k)}_{r3}$ for loops in Class III,
 which are partially $\E$cc-paired.
 Clearly, each loop $\kappa$ in Class III may be obtained from a loop in Class II
 by imposing $\E$cc-pair relations, which resulting in terms of the form of $|H^{IA}_{ij}|^2$.
 Below, we show that $F^{(k)}_{r3}$ has the same scaling behavior
 as $F^{(k)}_{r2}$ in Eq.(\ref{Fr2-scale}), i.e., $F^{(k)}_{r3} \sim d_\E^{-1}$.

 Let us first consider the case in which $\kappa$ contains only one $\E$cc pair.
 As see in  Eq.(\ref{il-il'}), this $\E$cc pair involves two steps: $s_{l} \to s_{l+1}$ and $s_{l'} \to s_{l'+1}$.
 Without the loss of generality, we assume that $l< l'$ and the step of $s_{l} \to s_{l+1}$ lies
 within the first half of the loop $\kappa$.
 (The opposite situation may be treated in a similar way.)

 Suppose that the step of $s_{l'} \to s_{l'+1}$ lies within the second half of $\kappa$.
 In the case of $l \ne 0$,
 it is not difficult to find that such loops give the following contribution to $F^{(k)}_{r3}$,
\begin{align}\label{F3-Q4-1}
  \sum_{i_l, i_{l+1}} Q_{s_0,s_{l}} Q_{s_{l+1},s_k} Q_{s_k, s_{l'}} Q_{s_{l'}, s_{2k}}
 |H^{IA}_{i_l i_{l+1}}|^2,
\end{align}
 where $Q_{s_{m},s_{n}}$ is defined as follows,
\begin{align}
 Q_{s_{m},s_{n}} = Q_{\gamma_{\rm sub}} \quad
 \text{ of $ \gamma_{\rm sub} = \{  i_{m+1}, \ldots, i_{n -1} \} $}.
\end{align}
 Making use of AR6, it is then direct to check that the above contribution scales as $d_\E^{-1}$.
 In the case of $l=0$,  one finds the following contribution to $F_{r3}^{(k)}$,
\begin{equation}
  \sum_{i_1} |H^{IA}_{i_0 i_{1}}|^2 F_{r2}^{(k-1)}\sim F_{r2}^{(k-1)} ,
\end{equation}
 possessing the same scaling behavior as discussed above [Eq.(\ref{Fr2-scale-app})].

 Similarly,  for $s_{l'} \to s_{l'+1}$ lying within the first half of $\kappa$,
 with a little more derivations, one finds the same scaling behavior of $d_\E^{-1}$.
 Therefore, the contribution of the case of only one $\E cc$-pair
 to $F^{(k)}_{r3}$ scales as $d_\E^{-1}$.

 Next, we discuss the cases of  more-than-one $\E cc$-pairs.
 Making use of results obtained above, these cases may be dealt with
 by the method of recursion.
 The result is similar, i.e., their contributions also scale as $ d_\E^{-1}$.
 Summarizing the above discussions, one sees that $F^{(k)}_{r3}$ scales as $d_\E^{-1}$.

 Further, similar to the case of $F^{(k)}_{r2}$ discussed above,
 signs of $F^{(k)}_{r3}$ should also have a random-type behavior.
 As a consequence, when a summation is taken over $r$ for getting $A_k$,
 its contribution is also negligible compared with that of $F^{(k)}_{r1}$.
 Finally, with $(F^{(k)}_{r2} + F^{(k)}_{r3})$ negligible,
 we reach Eq.(\ref{psir-sim-Fkr1}).

\section{Derivation of Eq.(\ref{Ykp-expression})}\label{app-compute-Y-lambda}

 In this appendix, we derive the expression of $Y^{(k)}$ in Eq.(\ref{Ykp-expression}).
 Let us write Eq.(\ref{Y-lambda2}) in the following form,
\begin{align}\label{}
 Y^{(k)} = \prod_{\substack{l=1 }}^{k-1} Y_{l},
\end{align}
 where
 \begin{align}
 \label{Y-l}
 Y_{l} =  {\sum_{\substack{i_l}} }' \frac{(\rho^\E_{\rm dos}(e^\E_{i_{l}}))^{-1}}{(E_n -E^0_{s_{l}}) (E_n -E^0_{s_{2k-l}})}.
 \end{align}
 For $s_{l} = (\alpha_l,i_l)$ and $s_{2k-l} = (\alpha_{2k-l} ,i_{2k-l})$
 belonging to one $\E$cc-pair and satisfying Eq.(\ref{p0:il=i2k-l}),
 $Y_l$ takes the following form,
\begin{equation}\label{Yl-sum-i}
	Y_l={\sum_{i}}'\frac{(\rho^\E_{\rm dos}(e^\E_{i}))^{-1}}
	{E_n-e^S_{\alpha}-e^\E_{i}}\frac{1}{E_n-e^S_{\beta}-e^\E_{i}}.
\end{equation}
 where subscripts ``$l$'' on the rhs are not written explicitly, for brevity.

 Let us compute the rhs of Eq.(\ref{Yl-sum-i}).
 In the case of $e^S_\alpha \neq e^S_\beta$, without loss of generality,
 let us consider $e^S_\alpha < e^S_\beta$.
 Due to largeness of the environmental density of states,
 the summation over the $i$-index may be approximated
 by an integration over the environmental energy, as shown in Eq.(\ref{sum-integral}).

 In the case of $(e^S_\beta - e^S_\alpha) > \Delta_{12}$,
 direct computation gives that
\begin{widetext}
\begin{align} \label{abn}
&{\sum_{i}}' \frac{(\rho^\E_{\rm dos}(e^\E_{i}))^{-1}}{(E_n-e_{\alpha}^S-e^\E_{i})(E_n-e_{\beta}^S-e^\E_{i})}\notag
\\
\simeq & \int_{e^\E_{\rm LB}}^{E^0_{r_1}-e^S_{\beta}}+\int^{{E^0_{r_1}-e^S_{\alpha}}}_{E^0_{r_2}-e^S_{\beta}}
+\int^{e^\E_{\rm UB}}_{E^0_{r_2}-e^S_{\alpha}}
\frac{  \mathrm{d}e^\E}{(E_n-e_{\alpha}^S-e^\E)(E_n-e_{\beta}^S-e^\E)}\notag
\\
=&\int_{e^\E_{\rm LB}}^{E^0_{r_1}-e^S_{\beta}}+\int^{{E^0_{r_1}-e^S_{\alpha}}}_{E^0_{r_2}-e^S_{\beta}}
+\int^{e^\E_{\rm UB}}_{E^0_{r_2}-e^S_{\alpha}}  \frac{1}{e^S_{\beta}-e^S_{\alpha}}\mathrm{d}
\ln\left|\frac{e^\E+e^S_{\alpha}-E_n}{e^\E+e^S_{\beta}-E_n} \right|  \notag
\\
=&\frac{1}{e_{\beta}^S-e_{\alpha}^S}(\ln\left|\frac{E^0_{r_1}-E_n
+e_{\alpha}^S-e_{\beta}^S}{E^0_{r_1}-E_n-e_{\alpha}^S
+e_{\beta}^S}\right|+\ln\left|\frac{E^0_{r_2}-E_n-e_{\alpha}^S+e_{\beta}^S
}{E^0_{r_2}-E_n+e_{\alpha}^S-e_{\beta}^S}\right|\notag\\+&\ln\left|\frac{
e^\E_{\rm UB}+e^S_{\alpha}-E_n}{e^\E_{\rm UB}+e^S_{\beta}-E_n}\right|-\ln\left|\frac{
e^\E_{\rm LB}+e^S_{\alpha}-E_n}{e^\E_{\rm LB}+e^S_{\beta}-E_n}\right| )\notag\\
\approx&\frac{2}{e_{\beta}^S-e_{\alpha}^S} \ln
\left(1+\frac{\Delta_{12}}{e_{\beta}^S-e_{\alpha}^S-\frac{\Delta_{12}}2} \right),\notag\\
\end{align}
where $\Delta_{12}$ is the energy scale of the set $\ov S_n$ as defined in Eq.(\ref{Delta-12}).
If $(e^S_\beta - e^S_\alpha)\gg\Delta_{12}$, the result can be further approximated as
$\dfrac{2 \Delta_{12}}{(e_{\beta}^S-e_{\alpha}^S)^2}$.

In the case of $(e^S_\beta - e^S_\alpha) < \Delta_{12}$, one finds that
  \begin{equation}
    \begin{aligned}
      &{\sum_{i}}' \frac{(\rho^\E_{\rm dos}(e^\E_{i}))^{-1}}{(E_n-e_{\alpha}^S-e^\E_{i})(E_n-e_{\beta}^S-e^\E_{i})}\\
      \approx &\int_{e^\E_{\rm LB}}^{E^0_{r_1}-e^S_{\beta}}+\int^{e^\E_{\rm UB}}_{E^0_{r_2}-e^S_\alpha}
      \frac{ \mathrm{d}e^\E}{(E_n-e_{\alpha}^S-e^\E)(E_n-e_{\beta}^S-e^\E)},\\
      =&\frac{1}{e_{\beta}^S-e_{\alpha}^S}
      \left(\ln\abs{\frac{E^0_{r_1}-e^S_{\beta}+e^S_{\alpha}-E_n}{E^0_{r_1}-E_n}}
      -\ln\abs{\frac{E^0_{r_2}-E_n}{E^0_{r_2}-e^S_{\alpha}+e^S_{\beta}-E_n}}\right.\\
      &\left.+\ln\abs{\frac{e^\E_{\rm UB}+e^S_{\alpha}-E_n}{e^\E_{\rm UB}+e^S_{\beta}-E_n}}
      -\ln\abs{\frac{e^\E_{\rm LB}+e^S_{\alpha}-E_n}{e^\E_{\rm LB}+e^S_{\beta}-E_n}}\right)\\
      \approx &\frac{2}{e_{\beta}^S-e_{\alpha}^S}\ln\left(1+2\frac{e^S_{\beta}-e^S_{\alpha}}{\Delta_{12}}\right)
    \end{aligned}
  \end{equation}
\color{black}

 In the case of $e^S_\alpha=e^S_\beta$, one finds that
\begin{align} \notag
 &{\sum_{i}}' \frac{(\rho^\E_{\rm dos}(e^\E_{i}))^{-1}}{(E_n-e_{\alpha}^S-e^\E_{i})^2}
 \approx \int_{e^\E_{\rm LB}}^{E_{r_1}-e^S_{\alpha}}
 +\int^{e^\E_{\rm UB}}_{E_{r_2}-e^S_{\alpha}}\frac{
\mathrm{d}e^\E}{(E_n-e_{\alpha}^S-e^\E)^2}\notag
 \\
=& \left( \frac{1}{E_{r_2}-E_n}-\frac{1}{E_{r_1}-E_n} \right.
 \left. +\frac{1}{E_n-e^S_{\alpha}-e^\E_{\rm UB}}-\frac{1}{E_n-e^S_{\alpha}-e^{\E}_{\rm LB}}  \right) 
\approx  \frac{4}{\Delta_{12}}. \label{abe}
\end{align}

 Thus, we get that
\begin{align}\label{Yl-final-result-nlam2}
 Y_l \approx \frac{4}{\Delta_{12}} K_{\alpha \beta}
\end{align}
 where $K_{\alpha \beta}$ is defined in Eq.(\ref{abe3}).
 Making use of Eq.(\ref{Yl-final-result-nlam2}),  it is straightforward to get Eq.(\ref{Ykp-expression}).
\end{widetext}

\section{Numerical details on the calculation of NPT region}\label{sect-ct-npt}
In this section, we give some numerical details we employ for the calculation of the NPT region.
To this end, let us recall Eq.~\eqref{nP-expan},
\begin{equation}
    |n_{P}\rangle= \sum_{l=1}^{k-1} \left( T_n \right)^{l} |n_Q  \rangle
 + \left( T_n \right)^k |n_P  \rangle .
\end{equation}
Introducing 
\begin{equation}
    \widetilde{A}_{k}=\langle n_{P}|(T_{n}^{\dagger})^{k}(T_{n})^{k}|n_{P}\rangle ,
\end{equation}
a convergent perturbation expansion of $|n_{P}\rangle$ exists if 
\begin{equation}
\lim_{k\rightarrow\infty}\widetilde{A}_{k}=0\ .
\end{equation}
Practically, in our numerical simulations we use the following criterion,
\begin{equation}
    \widetilde{A}_{k_{\text{max}}} \le \epsilon_{c},
\end{equation}
where we choose $k_{\text{max}} = 100$ and $\epsilon_{c} = 10^{-6}$. It should be noted that the results are not sensitive to these parameters.

\bibliographystyle{apsrev4-2-titles}
\bibliography{ref.bib}

@article{GORM,
  title = {Spin relaxation in a complex environment},
  author = {Esposito, Massimiliano and Gaspard, Pierre},
  journal = {Phys. Rev. E},
  volume = {68},
  issue = {6},
  pages = {066113},
  numpages = {21},
  year = {2003},
  month = {Dec},
  publisher = {American Physical Society},
  url = {https://link.aps.org/doi/10.1103/PhysRevE.68.066113}
}

@article{Genway13,
  title = {Dynamics of Thermalization and Decoherence of a Nanoscale System},
  author = {Genway, S. and Ho, A. F. and Lee, D. K. K.},
  journal = {Phys. Rev. Lett.},
  volume = {111},
  issue = {13},
  pages = {130408},
  numpages = {5},
  year = {2013},
  month = {Sep},
  publisher = {American Physical Society},
  url = {https://link.aps.org/doi/10.1103/PhysRevLett.111.130408}
}

@article{Berman07,
  title = {Decoherence and Thermalization},
  author = {Merkli, M. and Sigal, I. M. and Berman, G. P.},
  journal = {Phys. Rev. Lett.},
  volume = {98},
  issue = {13},
  pages = {130401},
  numpages = {4},
  year = {2007},
  month = {Mar},
  publisher = {American Physical Society},
  doi = {10.1103/PhysRevLett.98.130401},
  url = {https://link.aps.org/doi/10.1103/PhysRevLett.98.130401}
}

@article{Berman08,
title = {Dynamics of collective decoherence and thermalization},
journal = {Annals of Physics},
volume = {323},
number = {12},
pages = {3091-3112},
year = {2008},
issn = {0003-4916},
doi = {https://doi.org/10.1016/j.aop.2008.07.004},
url = {https://www.sciencedirect.com/science/article/pii/S0003491608001140},
author = {M. Merkli and G.P. Berman and I.M. Sigal}
}

@article{Gorin14,
  title = {Single-qubit decoherence under a separable coupling to a random matrix environment},
  author = {Carrera, M. and Gorin, T. and Seligman, T. H.},
  journal = {Phys. Rev. A},
  volume = {90},
  issue = {2},
  pages = {022107},
  numpages = {12},
  year = {2014},
  month = {Aug},
  publisher = {American Physical Society},
  doi = {10.1103/PhysRevA.90.022107},
  url = {https://link.aps.org/doi/10.1103/PhysRevA.90.022107}
}

@article{Jin13,
  title = {Quantum decoherence scaling with bath size: Importance of dynamics, connectivity, and randomness},
  author = {Jin, Fengping and Michielsen, Kristel and Novotny, Mark A. and Miyashita, Seiji and Yuan, Shengjun and De Raedt, Hans},
  journal = {Phys. Rev. A},
  volume = {87},
  issue = {2},
  pages = {022117},
  numpages = {13},
  year = {2013},
  month = {Feb},
  publisher = {American Physical Society},
  url = {https://link.aps.org/doi/10.1103/PhysRevA.87.022117}
}

@article{Olshanii14,
  title = {Typical, finite baths as a means of exact simulation of open quantum systems},
  author = {Silvestri, Luciano and Jacobs, Kurt and Dunjko, Vanja and Olshanii, Maxim},
  journal = {Phys. Rev. E},
  volume = {89},
  issue = {4},
  pages = {042131},
  numpages = {9},
  year = {2014},
  month = {Apr},
  publisher = {American Physical Society},
  url = {https://link.aps.org/doi/10.1103/PhysRevE.89.042131}
}

@article{Jin17,
  title = {Relaxation, thermalization, and Markovian dynamics of two spins coupled to a spin bath},
  author = {De Raedt, H. and Jin, F. and Katsnelson, M. I. and Michielsen, K.},
  journal = {Phys. Rev. E},
  volume = {96},
  issue = {5},
  pages = {053306},
  numpages = {17},
  year = {2017},
  month = {Nov},
  publisher = {American Physical Society},
  url = {https://link.aps.org/doi/10.1103/PhysRevE.96.053306}
}

@article{Yuan11,
  title={Decoherence and thermalization of quantum spin systems},
  author={Yuan, Shengjun},
  journal={Journal of Computational and Theoretical Nanoscience},
  volume={8},
  number={6},
  pages={889--911},
  year={2011},
  publisher={American Scientific Publishers}
}

@article{Bulaev05,
  title = {Spin Relaxation and Decoherence of Holes in Quantum Dots},
  author = {Bulaev, Denis V. and Loss, Daniel},
  journal = {Phys. Rev. Lett.},
  volume = {95},
  issue = {7},
  pages = {076805},
  numpages = {4},
  year = {2005},
  month = {Aug},
  publisher = {American Physical Society},
  url = {https://link.aps.org/doi/10.1103/PhysRevLett.95.076805}
}

@article{Fialko15,
  title = {Quantum heat baths satisfying the eigenstate thermalization hypothesis},
  author = {Fialko, O.},
  journal = {Phys. Rev. E},
  volume = {92},
  issue = {2},
  pages = {022104},
  numpages = {5},
  year = {2015},
  month = {Aug},
  publisher = {American Physical Society},
  url = {https://link.aps.org/doi/10.1103/PhysRevE.92.022104}
}

@article{Wang08,
  title = {Entanglement-induced decoherence and energy eigenstates},
  author = {Wang, Wen-ge and Gong, Jiangbin and Casati, G. and Li, Baowen},
  journal = {Phys. Rev. A},
  volume = {77},
  issue = {1},
  pages = {012108},
  numpages = {4},
  year = {2008},
  month = {Jan},
  publisher = {American Physical Society},
  url = {https://link.aps.org/doi/10.1103/PhysRevA.77.012108}
}

@article{Prosen181,
  title = {Many-Body Quantum Chaos: Analytic Connection to Random Matrix Theory},
  author = {Kos, Pavel and Ljubotina, Marko and Prosen, Toma\ifmmode \check{z}\else \v{z}\fi{}},
  journal = {Phys. Rev. X},
  volume = {8},
  issue = {2},
  pages = {021062},
  numpages = {11},
  year = {2018},
  month = {Jun},
  publisher = {American Physical Society},
  url = {https://link.aps.org/doi/10.1103/PhysRevX.8.021062}
}

@article{Prosen182,
  title = {Exact Spectral Form Factor in a Minimal Model of Many-Body Quantum Chaos},
  author = {Bertini, Bruno and Kos, Pavel and Prosen, Toma\ifmmode \check{z}\else \v{z}\fi{}},
  journal = {Phys. Rev. Lett.},
  volume = {121},
  issue = {26},
  pages = {264101},
  numpages = {6},
  year = {2018},
  month = {Dec},
  publisher = {American Physical Society},
  url = {https://link.aps.org/doi/10.1103/PhysRevLett.121.264101}
}

@article{Berry77,
doi = {10.1088/0305-4470/10/12/016},
url = {https://doi.org/10.1088/0305-4470/10/12/016},
year = {1977},
month = {dec},
publisher = {},
volume = {10},
number = {12},
pages = {2083},
author = {M V Berry},
title = {Regular and irregular semiclassical wavefunctions},
journal = {Journal of Physics A: Mathematical and General} 
}

@article{KpHl,
  title = {Bound-State Eigenfunctions of Classically Chaotic Hamiltonian Systems: Scars of Periodic Orbits},
  author = {Heller, Eric J.},
  journal = {Phys. Rev. Lett.},
  volume = {53},
  issue = {16},
  pages = {1515--1518},
  numpages = {0},
  year = {1984},
  month = {Oct},
  publisher = {American Physical Society},
  url = {https://link.aps.org/doi/10.1103/PhysRevLett.53.1515}
}

@article{Heller87,
  title = {Properties of random superpositions of plane waves},
  author = {O'Connor, P. and Gehlen, J. and Heller, E. J.},
  journal = {Phys. Rev. Lett.},
  volume = {58},
  issue = {13},
  pages = {1296--1299},
  numpages = {0},
  year = {1987},
  month = {Mar},
  publisher = {American Physical Society},
  doi = {10.1103/PhysRevLett.58.1296},
  url = {https://link.aps.org/doi/10.1103/PhysRevLett.58.1296}
}

@article{Sr96,
  title = {Gaussian random eigenfunctions and spatial correlations in quantum dots},
  author = {Srednicki, Mark},
  journal = {Phys. Rev. E},
  volume = {54},
  issue = {1},
  pages = {954--955},
  numpages = {0},
  year = {1996},
  month = {Jul},
  publisher = {American Physical Society},
  url = {https://link.aps.org/doi/10.1103/PhysRevE.54.954}
}

@article{Sr98,
  title = {Random matrix elements and eigenfunctions in chaotic systems},
  author = {Hortikar, Sanjay and Srednicki, Mark},
  journal = {Phys. Rev. E},
  volume = {57},
  issue = {6},
  pages = {7313--7316},
  numpages = {0},
  year = {1998},
  month = {Jun},
  publisher = {American Physical Society},
  url = {https://link.aps.org/doi/10.1103/PhysRevE.57.7313}
}

@article{Bies01,
  title = {Localization of eigenfunctions in the stadium billiard},
  author = {Bies, W. E. and Kaplan, L. and Haggerty, M. R. and Heller, E. J.},
  journal = {Phys. Rev. E},
  volume = {63},
  issue = {6},
  pages = {066214},
  numpages = {16},
  year = {2001},
  month = {May},
  publisher = {American Physical Society},
  doi = {10.1103/PhysRevE.63.066214},
  url = {https://link.aps.org/doi/10.1103/PhysRevE.63.066214}
}

@article{Backer02,
doi = {10.1088/0305-4470/35/3/307},
url = {https://doi.org/10.1088/0305-4470/35/3/307},
year = {2002},
month = {jan},
publisher = {},
volume = {35},
number = {3},
pages = {539},
author = {Arnd Bäcker and Roman Schubert},
title = {Autocorrelation function of eigenstates in chaotic and mixed systems},
journal = {Journal of Physics A: Mathematical and General}
}

@article{Urb03,
doi = {10.1088/0305-4470/36/38/102},
url = {https://doi.org/10.1088/0305-4470/36/38/102},
year = {2003},
month = {sep},
publisher = {},
volume = {36},
number = {38},
pages = {L495},
author = {Juan Diego Urbina and Klaus Richter},
title = {Supporting random wave models: a quantum mechanical approach},
journal = {Journal of Physics A: Mathematical and General}
}

@article{Urb04,
  title = {Semiclassical construction of random wave functions for confined systems},
  author = {Urbina, Juan Diego and Richter, Klaus},
  journal = {Phys. Rev. E},
  volume = {70},
  issue = {1},
  pages = {015201},
  numpages = {4},
  year = {2004},
  month = {Jul},
  publisher = {American Physical Society},
  url = {https://link.aps.org/doi/10.1103/PhysRevE.70.015201}
}

@article{Urb06,
  title = {Statistical Description of Eigenfunctions in Chaotic and Weakly Disordered Systems beyond Universality},
  author = {Urbina, Juan Diego and Richter, Klaus},
  journal = {Phys. Rev. Lett.},
  volume = {97},
  issue = {21},
  pages = {214101},
  numpages = {4},
  year = {2006},
  month = {Nov},
  publisher = {American Physical Society},
  url = {https://link.aps.org/doi/10.1103/PhysRevLett.97.214101}
}

@article{Kp05,
  title = {Correlation function bootstrapping in quantum chaotic systems},
  author = {Kaplan, L.},
  journal = {Phys. Rev. E},
  volume = {71},
  issue = {5},
  pages = {056212},
  numpages = {12},
  year = {2005},
  month = {May},
  publisher = {American Physical Society},
  url = {https://link.aps.org/doi/10.1103/PhysRevE.71.056212}
}

@article{pre18-EF-BC,
  title = {Characterization of random features of chaotic eigenfunctions in unperturbed basis},
  author = {Wang, Jiaozi and Wang, Wen-ge},
  journal = {Phys. Rev. E},
  volume = {97},
  issue = {6},
  pages = {062219},
  numpages = {11},
  year = {2018},
  month = {Jun},
  publisher = {American Physical Society},
  url = {https://link.aps.org/doi/10.1103/PhysRevE.97.062219}
}

@article{Meredith88,
  title = {Quantum chaos in a schematic shell model},
  author = {Meredith, D. C. and Koonin, S. E. and Zirnbauer, M. R.},
  journal = {Phys. Rev. A},
  volume = {37},
  issue = {9},
  pages = {3499--3513},
  numpages = {0},
  year = {1988},
  month = {May},
  publisher = {American Physical Society},
  url = {https://link.aps.org/doi/10.1103/PhysRevA.37.3499}
}

@article{EFchaos-Benet03,
doi = {10.1088/0305-4470/36/5/307},
url = {https://doi.org/10.1088/0305-4470/36/5/307},
year = {2003},
month = {jan},
publisher = {},
volume = {36},
number = {5},
pages = {1289},
author = {L Benet and J Flores and H Hern\'andez-Salda\~na
 and F M Izrailev and F Leyvraz and T H Seligman},
title = {Fluctuations of wavefunctions about their classical average},
journal = {Journal of Physics A: Mathematical and General}
}

@article{EFchaos-Borgonovi98,
  title = {Quantum-classical correspondence in energy space: Two interacting spin particles},
  author = {Borgonovi, F. and Guarneri, I. and Izrailev, F. M.},
  journal = {Phys. Rev. E},
  volume = {57},
  issue = {5},
  pages = {5291--5302},
  numpages = {0},
  year = {1998},
  month = {May},
  publisher = {American Physical Society},
  doi = {10.1103/PhysRevE.57.5291},
  url = {https://link.aps.org/doi/10.1103/PhysRevE.57.5291}
}

@article{EFchaos-Benet00,
title = {Semiclassical properties of eigenfunctions and occupation number distribution for a model of two interacting particles},
journal = {Physics Letters A},
volume = {277},
number = {2},
pages = {87-93},
year = {2000},
issn = {0375-9601},
doi = {https://doi.org/10.1016/S0375-9601(00)00692-7},
url = {https://www.sciencedirect.com/science/article/pii/S0375960100006927},
author = {L. Benet and F.M. Izrailev and T.H. Seligman and A. Suárez-Moreno}
}

@article{Deutch91,
  title = {Quantum statistical mechanics in a closed system},
  author = {Deutsch, J. M.},
  journal = {Phys. Rev. A},
  volume = {43},
  issue = {4},
  pages = {2046--2049},
  numpages = {0},
  year = {1991},
  month = {Feb},
  publisher = {American Physical Society},
  doi = {10.1103/PhysRevA.43.2046},
  url = {https://link.aps.org/doi/10.1103/PhysRevA.43.2046}
}

@article{srednicki1994chaos,
  title = {Chaos and quantum thermalization},
  author = {Srednicki, Mark},
  journal = {Phys. Rev. E},
  volume = {50},
  issue = {2},
  pages = {888--901},
  numpages = {0},
  year = {1994},
  month = {Aug},
  publisher = {American Physical Society},
  doi = {10.1103/PhysRevE.50.888},
  url = {https://link.aps.org/doi/10.1103/PhysRevE.50.888}
}

@article{srednicki-JPA96,
doi = {10.1088/0305-4470/29/4/003},
url = {https://doi.org/10.1088/0305-4470/29/4/003},
year = {1996},
month = {feb},
publisher = {},
volume = {29},
number = {4},
pages = {L75},
author = {Mark Srednicki},
title = {Thermal fluctuations in quantized chaotic systems},
journal = {Journal of Physics A: Mathematical and General}
}

@article{srednicki1999approach,
doi = {10.1088/0305-4470/32/7/007},
url = {https://doi.org/10.1088/0305-4470/32/7/007},
year = {1999},
month = {feb},
publisher = {},
volume = {32},
number = {7},
pages = {1163},
author = {Mark Srednicki},
title = {The approach to thermal equilibrium in quantized chaotic systems},
journal = {Journal of Physics A: Mathematical and General}
}

@article{RS-PRL12,
  title = {Alternatives to Eigenstate Thermalization},
  author = {Rigol, Marcos and Srednicki, Mark},
  journal = {Phys. Rev. Lett.},
  volume = {108},
  issue = {11},
  pages = {110601},
  numpages = {5},
  year = {2012},
  month = {Mar},
  publisher = {American Physical Society},
  url = {https://link.aps.org/doi/10.1103/PhysRevLett.108.110601}
}

@article{Rigol-Aip2016,
author = {Luca D'Alessio and Yariv Kafri and Anatoli Polkovnikov and Marcos Rigol},
title = {From quantum chaos and eigenstate thermalization to statistical mechanics and thermodynamics},
journal = {Advances in Physics},
volume = {65},
number = {3},
pages = {239--362},
year = {2016},
publisher = {Taylor \& Francis},
doi = {10.1080/00018732.2016.1198134},
URL = {https://doi.org/10.1080/00018732.2016.1198134
}
}

@article{Deutch-RPP18,
doi = {10.1088/1361-6633/aac9f1},
url = {https://doi.org/10.1088/1361-6633/aac9f1},
year = {2018},
month = {jul},
publisher = {IOP Publishing},
volume = {81},
number = {8},
pages = {082001},
author = {Deutsch, Joshua M},
title = {Eigenstate thermalization hypothesis},
journal = {Reports on Progress in Physics}
}

@article{pre25-ETHsc,
  title = {Semiclassical study of diagonal and off-diagonal functions in the eigenstate thermalization hypothesis},
  author = {Wang, Xiao and Wang, Wen-ge},
  journal = {Phys. Rev. E},
  volume = {112},
  issue = {5},
  pages = {054215},
  numpages = {14},
  year = {2025},
  month = {Nov},
  publisher = {American Physical Society},
  url = {https://link.aps.org/doi/10.1103/xbtc-hlxw}
}

@article{PhysRevLett.134.010404-RMT-EF,
  title = {Universal Correlations in Chaotic Many-Body Quantum States: Fock-Space Formulation of Berry's Random Wave Model},
  author = {Schoeppl, Florian and Dubertrand, R\'emy and Urbina, Juan-Diego and Richter, Klaus},
  journal = {Phys. Rev. Lett.},
  volume = {134},
  issue = {1},
  pages = {010404},
  numpages = {6},
  year = {2025},
  month = {Jan},
  publisher = {American Physical Society},
  doi = {10.1103/PhysRevLett.134.010404},
  url = {https://link.aps.org/doi/10.1103/PhysRevLett.134.010404}
}

@misc{WW-ETH-conf,
      title={An operator-Weyl-symbol approach to eigenstate thermalization hypothesis}, 
      author={Wang, Xiao Wang and Wang, Wen-ge},
      year={2025},
      eprint={2509.24490},
      archivePrefix={arXiv},
      primaryClass={quant-ph},
      url={https://arxiv.org/abs/2509.24490}, 
}

@article{WIC98,
  title = {Structure of eigenstates and local spectral density of states: A three-orbital schematic shell model},
  author = {Wang, Wen-ge and Izrailev, F. M. and Casati, G.},
  journal = {Phys. Rev. E},
  volume = {57},
  issue = {1},
  pages = {323--339},
  numpages = {0},
  year = {1998},
  month = {Jan},
  publisher = {American Physical Society},
  url = {https://link.aps.org/doi/10.1103/PhysRevE.57.323}
}

@article{pre02-GBW,
  title = {Nonperturbative and perturbative parts of energy eigenfunctions: A three-orbital schematic shell model},
  author = {Wang, Wen-ge},
  journal = {Phys. Rev. E},
  volume = {65},
  issue = {3},
  pages = {036219},
  numpages = {9},
  year = {2002},
  month = {Feb},
  publisher = {American Physical Society},
  url = {https://link.aps.org/doi/10.1103/PhysRevE.65.036219}
}

@article{pre00-GBW,
  title = {Perturbative and nonperturbative parts of eigenstates and local spectral density of states: The Wigner-band random-matrix model},
  author = {Wang, Wen-ge},
  journal = {Phys. Rev. E},
  volume = {61},
  issue = {1},
  pages = {952--955},
  numpages = {0},
  year = {2000},
  month = {Jan},
  publisher = {American Physical Society},
  doi = {10.1103/PhysRevE.61.952},
  url = {https://link.aps.org/doi/10.1103/PhysRevE.61.952}
}

@article{CPL04,
title = {Decay Rate of Energy Eigenfunctions in Classically Energetically Inaccessible Regions},
journal = {Chin. Phys. Lett.},
volume = {21},
number = {10},
pages = {1869-1872},
year = {2004},
issn = {},
url = {http://cpl.iphy.ac.cn/en/article/id/37175},
author = {Wang Wen-ge}
}

@article{CPL05,
doi = {10.1088/0256-307X/22/12/002},
url = {https://doi.org/10.1088/0256-307X/22/12/002},
year = {2005},
month = {dec},
publisher = {},
volume = {22},
number = {12},
pages = {2991},
author = {Wang Wen-Ge},
title = {Decay Rate of Energy Eigenfunctions in Classically
Energetically Inaccessible Regions in more than One-Dimensional
Configuration Spaces},
journal = {Chinese Physics Letters}
}

@article{JPA19-NPT-C,
  title={Convergent perturbation expansion of energy eigenfunctions on unperturbed basis states in classically-forbidden regions},
  author={Wang, Jiaozi and Wang, Wen-ge},
  journal={Journal of Physics A: Mathematical and Theoretical},
  volume={52},
  number={23},
  pages={235204},
  year={2019},
url = {https://doi.org/10.1088/1751-8121/ab1c07},
  publisher={IOP Publishing}
}

@article{CTP19-renorm,
doi = {10.1088/0253-6102/71/7/861},
url = {https://doi.org/10.1088/0253-6102/71/7/861},
year = {2019},
month = {jul},
publisher = {Chinese Physical Society and IOP Publishing Ltd},
volume = {71},
number = {7},
pages = {861},
author = {Wang, Wen-Ge},
title = {A Renormalized-Hamiltonian-Flow Approach to Eigenenergies and Eigenfunctions},
journal = {Communications in Theoretical Physics}
}

@article{pre20-mc-can,
  title = {Closeness of the reduced density matrix of an interacting small system to the Gibbs state},
  author = {Wang, Wen-ge},
  journal = {Phys. Rev. E},
  volume = {102},
  issue = {1},
  pages = {012127},
  numpages = {15},
  year = {2020},
  month = {Jul},
  publisher = {American Physical Society},
  doi = {10.1103/PhysRevE.102.012127},
  url = {https://link.aps.org/doi/10.1103/PhysRevE.102.012127}
}

@article{Nation_2018,
doi = {10.1088/1367-2630/aae28f},
url = {https://doi.org/10.1088/1367-2630/aae28f},
year = {2018},
month = {oct},
publisher = {IOP Publishing},
volume = {20},
number = {10},
pages = {103003},
author = {Nation, Charlie and Porras, Diego},
title = {Off-diagonal observable elements from random matrix theory: distributions, fluctuations, and eigenstate thermalization},
journal = {New Journal of Physics}
}

@article{PhysRevX.15.011059-Luca,
  title = {Hydrodynamics and the Eigenstate Thermalization Hypothesis},
  author = {Capizzi, Luca and Wang, Jiaozi and Xu, Xiansong and Mazza, Leonardo and Poletti, Dario},
  journal = {Phys. Rev. X},
  volume = {15},
  issue = {1},
  pages = {011059},
  numpages = {21},
  year = {2025},
  month = {Mar},
  publisher = {American Physical Society},
  doi = {10.1103/PhysRevX.15.011059},
  url = {https://link.aps.org/doi/10.1103/PhysRevX.15.011059}
}

@article{offeth_PhysRevE.61.R2180,
  title = {Trace formula for products of diagonal matrix elements in chaotic systems},
  author = {Hortikar, Sanjay and Srednicki, Mark},
  journal = {Phys. Rev. E},
  volume = {61},
  issue = {3},
  pages = {R2180--R2183},
  numpages = {0},
  year = {2000},
  month = {Mar},
  publisher = {American Physical Society},
  doi = {10.1103/PhysRevE.61.R2180},
  url = {https://link.aps.org/doi/10.1103/PhysRevE.61.R2180}
}

@article{offeth_Wilkinson_1987,
url = {https://dx.doi.org/10.1088/0305-4470/20/9/028},
year = {1987},
month = {jun},
publisher = {},
volume = {20},
number = {9},
pages = {2415},
author = {M Wilkinson},
title = {A semiclassical sum rule for matrix elements of classically chaotic systems},
journal = {Journal of Physics A: Mathematical and General}
}

\end{document}